\documentclass{revtex4}
 \pdfoutput=1
\usepackage{ae,aecompl}
\usepackage{helvet}

\usepackage[T1]{fontenc}
\usepackage[latin9]{inputenc}
\usepackage{amsmath}
\usepackage{amssymb}
\usepackage{graphicx}
\usepackage[unicode=true,pdfusetitle,
 bookmarks=true,bookmarksnumbered=false,bookmarksopen=false,
 breaklinks=false,pdfborder={0 0 1},backref=false,colorlinks=false]
 {hyperref}

\makeatletter
\@ifundefined{textcolor}{}
{%
 \definecolor{BLACK}{gray}{0}
 \definecolor{WHITE}{gray}{1}
 \definecolor{RED}{rgb}{1,0,0}
 \definecolor{GREEN}{rgb}{0,1,0}
 \definecolor{BLUE}{rgb}{0,0,1}
 \definecolor{CYAN}{cmyk}{1,0,0,0}
 \definecolor{MAGENTA}{cmyk}{0,1,0,0}
 \definecolor{YELLOW}{cmyk}{0,0,1,0}
}

\makeatother

\begin{document}

\title{Giant magnetoresistance in the variable range hopping regime}

\author{L. Ioffe}

\affiliation{LPTHE, Universite Pierre and Marie Curie, Boite 126, T13-14 4eme
etage, 4 place Jussieu, Paris CEDEX 05, France and Department of Physics,
Rutgers University 136 Frelinghuysen Rd, Piscataway, New Jersey 08854
USA}

\author{B.Spivak}

\affiliation{Department of Physics, University of Washington, Seattle WA 98195,
USA}
\begin{abstract}
We predict the universal power law dependence of localization length
on magnetic field in the strongly localized regime. This effect is
due to the orbital quantum interference. Physically, this dependence
shows up in an anomalously large negative magnetoresistance in the
hopping regime. The reason for the universality is that the problem
of the electron tunneling in a random media belongs to the same universality
class as directed polymer problem even in the case of wave functions
of random sign. We present numerical simulations which prove this
conjecture. We discuss the existing experiments that show anomalously
large magnetoresistance. We also discuss the role of localized spins
in real materials and the spin polarizing effect of magnetic field. 
\end{abstract}


\maketitle

\section{Introduction}

In strongly disordered conductors, single electrons states are localized,
so the conductivity is due to phonon assisted electron tunneling between
localized states. The length of a typical hop $r_{hop}$ grows as
temperature is decreased and becomes much larger than the distance
between the localized states in the variable range hopping regime.
\cite{Efros1985,Mott1990} In this paper we study the orbital mechanism
of the magnetoresistance in this regime. We show that at sufficiently
low temperatures it is due to the localization length dependence on
magnetic field, $B$, and that it is given by a universal power law.
This localization length dependence on magnetic field translates into
an exponentially large variation of the resistance. The sign of the
orbital magnetoresistance depends on the details of impurity scattering,
but in the typical case the low temperature magnetoresistance is negative.
Similar to the metallic regime, the origin of the negative magnetoresistance
is the electron quantum interference, however, the amplitudes that
interfere correspond to different processes in these two cases. Despite
its much larger magnitude the negative magnetoresistance in the hopping
regime received much less attention, both theoretically and experimentally,
than its counterpart in the metallic regime. One of goals of this
paper is to draw the attention of the community to this interesting
phenomenon. 

\begin{figure}[ptb]
\includegraphics[width=0.4\textwidth]{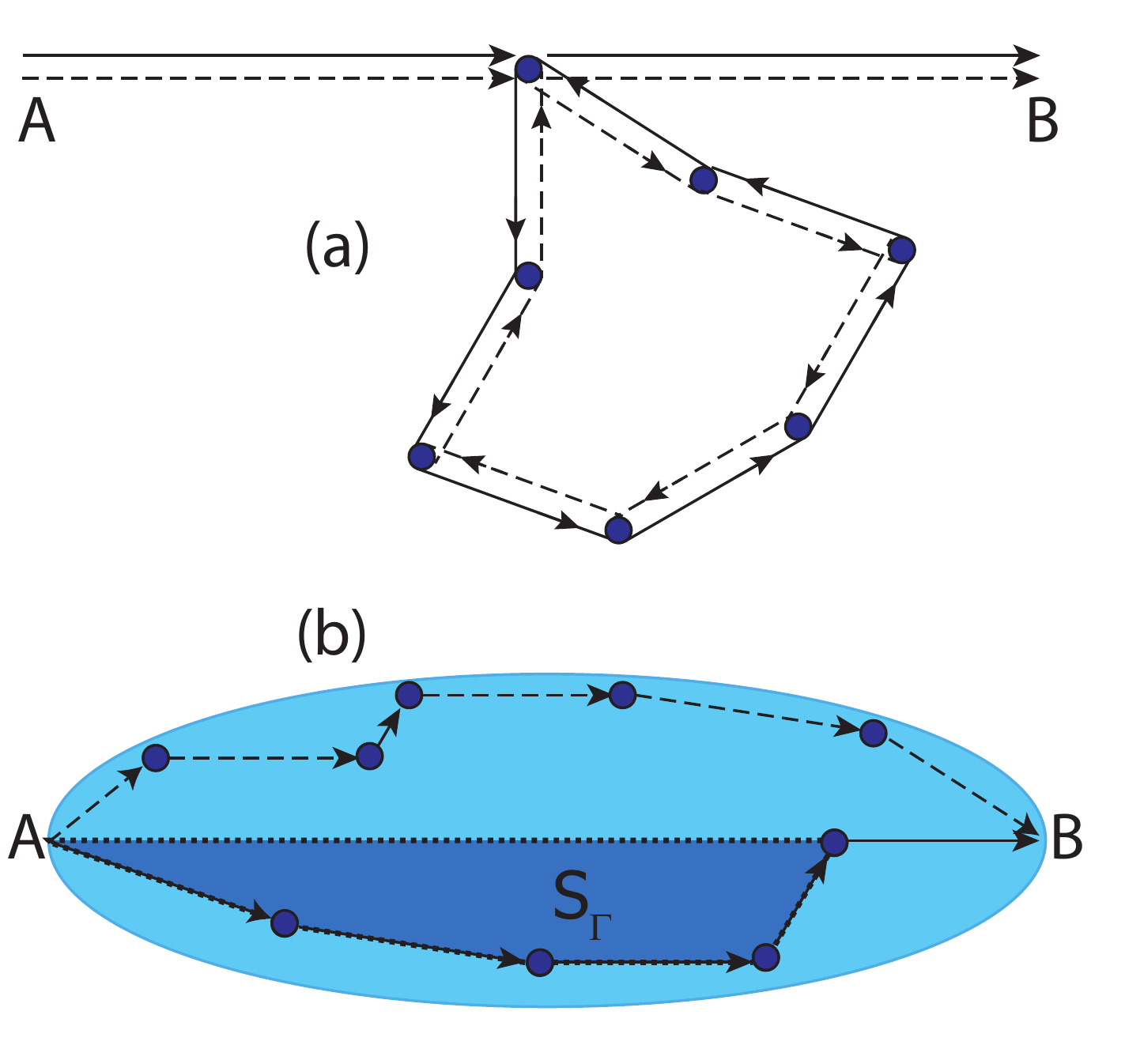} \caption{Qualitative picture of the interference effects in disordered metals.
Panel (a) shows interference in the weak localization regime that
is due to self-crossing diffusive paths. Quantum propagation from
A to B is the sum of two amplitudes that contain clockwise and counterclockwise
motion along the loop that is a part of the self intersecting path.
Panel (b) shows interference in hopping regime in which the backward
motion of electrons gives negligible contribution to the tunneling
between sites A and B. In this case typical paths contributing to
the interference are located in the shaded (light blue) area with
transverse direction that scales with the length of the hop, $L$.
The magnetic field has a significant effect if the flux trough the
area, $S_{\Gamma}$, formed by a typical path and a straight line
(dark blue) is of the order of one flux quantum. }

\label{fig:Interference} 
\end{figure}

We begin with a brief review of the nature of the magnetoresistance
in metals. The conventional theory of magnetoresistance associates
it with the classical effect of electron motion along cyclotron orbits.
For a typical metal the magnetoresistance is controlled by the parameter
$(\omega_{{\rm c}}\tau_{\textrm{tr}})^{2}$. Here $\omega_{c}$ is
the cyclotron frequency, and $\tau_{tr}$ is the transport mean free
time, (see e.g. \cite{Abrikosov1988}). In contrast to these expectations,
many disordered metals show negative magnetoresistance at small magnetic
fields. The negative magnetoresistance in weakly disordered metals
has been explained in the framework of the weak localization theory,
which takes into account the quantum interference of probability amplitudes
for electrons to travel along self-intersecting diffusive paths \cite{Altshuler1979,Hikami1980,Larkin1980,Lee1985}
such as shown in the Fig.~\ref{fig:Interference}a. The interfering
amplitudes correspond to the clockwise and counterclockwise propagation
of the electron wave along the loop formed by the self-intersecting
path. In the absence of magnetic field these amplitudes interfere
constructively increasing the probability of return to the intersection
point. In the presence of magnetic field these amplitudes acquire
different phases, and the interference is suppressed leading to the
negative magnetoresistance. The magnitude of negative magnetoresistance
in this regime is relatively small because it scales with the small
parameter $1/k_{F}l_{tr}$. Here $k_{F}$ is the Fermi momentum, and
$l_{tr}$ is the transport mean free path.

Experimentally, in many materials the magnetoresistance in the hopping
regime is significantly larger than in the metallic regime. A positive
magnetoresistance of several orders of magnitude in hopping regime
has been observed long ago (see, e.g. Ref.~\cite{Efros1985} and
references therein). Significant negative magnetoresistance in variable
range hopping regime ranging up to two orders of magnitude, has been
observed in many experimental works \cite{Savchenko1987,Jiang1992,Ovadyahu1990,Ovadyahu1995,Kravchenko1998,Spivak2010,Wang2006,Hong2011,Friedman1996,Mitin2007,Dynes1996}.
In some of these works a large anisotropy of the negative magnetoresistance
has been observed in 2D samples , indicating its orbital nature.

Phonon emission and absorption make different hopping events incoherent,
whilst the electron tunneling between the localized states is a quantum
mechanical process. The magnetoresistance is due to the magnetic field
dependence of the probability of one hop. Qualitatively, large orbital
magnetoresistance in the hopping regime is due the interference of
the tunneling amplitudes along different tunneling paths contributing
to a single hop that are distributed in a cigar-shaped region shown
in \ref{fig:Interference}b. In this regime the tunneling paths containing
loops give exponentially small contribution to the tunneling probability.
This is the main difference from the weak localization where the interference
is due to the paths that circle a loop (see \ref{fig:Interference}a).
Because in the variable range hopping regime electrons hop over distances
much larger than the distance between localized states, the cigar-shaped
region contains many electron scatterers. The amplitudes, $\mu_{i}$,
describing individual scattering process at state $i$ may be positive
and negative. The sign distribution of $\mu_{i}$ determines the sign
of the magnetoresistance, as we explain below in section \ref{sub:The-sign-transition.}. 

Large positive magnetoresistance may be associated with a shrinkage
of the hydrogen-like localized electron wave functions at the scales
less than the inter-impurity distance. Quantitatively this picture
works well only in a very high magnetic field and at sufficiently
high temperatures at which the typical electron hopping length is
shorter than the distance between impurities. A theory of the positive
magnetoresistance which takes into account the electron scattering
with positive scattering amplitudes has been developed in works \cite{Shklovskii1982,Shklovskii1983a,Shklovskii1983b,Shklovskii1983c,Shklovskii1984}.
In this case the tunneling amplitudes interfere constructively in
the absence of the field, while the phases induced by the magnetic
field destroy this interference.

An orbital mechanism of the negative magnetoresistance may be associated
with the randomness of the signs of the scattering amplitudes, $\mu_{i}$,
that is due to random sign of $\epsilon-\epsilon_{i}$ .\cite{Nguen1985,Shklovskii1990a,Medina1990,Shklovskii1991,Zhao1991,Nguen1986}
Here $\epsilon$ is the energy of the tunneling electron and $\epsilon_{i}$
is the energy of a localized state.. This sign randomness may lead
to random signs of the interfering tunneling amplitudes at $B=0$.
The magnetic field makes tunneling amplitudes complex which increases
the conductance in this situation. Thus, the sign of the orbital magnetoresistance
is related to the sign distribution of the localized electron wave
functions.

In this work we develop a quantitative theory of the orbital magnetoresistance
in the hopping regime and discuss the available experimental data
in the light of our results. Because most of experiments have been
done on two dimensional samples we will focus on the two dimensional
hopping regime of the electrons and corresponding experiments. 

We show that for physically relevant cases even a small concentration
of impurities with $\mu_{i}$<0 leads to completely random signs of
the tunneling amplitudes at large scales. Therefore, at sufficiently
low temperatures and small magnetic field the variable range hopping
magnetoresistance is negative. At higher magnetic field and higher
temperatures it can be both positive and negative.

The plan of the paper is as follows: In section \ref{sub:Review-of-variable}
we start with a brief review of the basis of variable range hopping
theory, and discuss qualitative picture of the variable range hopping
magnetoresistance. In sections \ref{sub:Statistics-of-the} , \ref{sub:The-sign-transition.}
we discuss the statistics of the modulus and of the sign of the localized
electron wave function. In particular, in section \ref{sub:The-sign-transition.}
we discuss the conditions for the existence of the ``sign phase transition''
where, as a function of the concentration of scatterers with $\mu_{i}$<0,
the system changes from the sign ordered to sign disordered phases.
In section \ref{sec:Magnetoresistance-in-the} we apply the theory
developed in section \ref{sec:Electron-propagation-in} to compute
the magnetoresistance. The section \ref{sec:Application-to-other}
discusses applications of the results for the sign phase transition
to other physical systems. Finally, section \ref{sec:Review-of-the}
gives a short review of the experimental situation.

\section{Electron transport in variable range hopping regime. \label{sec:Electron-propagation-in}}

\subsection{Review of variable range hopping theory.\label{sub:Review-of-variable} }

In the localized regime the electron wave functions decay exponentially
with the distance, $\mathbf{|r}-\mathbf{r}_{i}|,$ from the impurity:$\psi_{i}\left(\mathbf{r}\right)\sim\exp(-|\mathbf{r}-\mathbf{r}_{i}|/\xi)$
where $\mathbf{r}_{i}$ is the center of the localized wave function
and $\xi$ is a typical localization radius. In this case the conductivity
is determined by phonon assisted electron hopping between localized
states. At low temperatures the typical hopping length $r_{hop}$
is determined by the competition between two exponential factors:
the hopping probability $W_{ij}$ that decays exponentially with the
distance between impurities $r_{ij}$ and the thermal factor, $\exp(-E_{hop}(r_{ij})/T)$
where $E_{hop}(r_{ij})$ is the hopping activation energy that decreases
with $r_{ij}$. These factors give the exponential dependence of the
typical hopping rate at distance $r_{hop}$: $\exp(-E_{hop}(r)/T-2r/\xi)$.
This exponential factor is maximal for the typical hopping length,
$r_{hop}$, which is much larger than the distance between localized
states, as illustrated in Fig.~\ref{fig:Interference}b:
\begin{equation}
r_{hop}\sim\left(\frac{T_{0}}{T}\right)^{\zeta}\xi,\label{eq:r_VRH}
\end{equation}
As a result, the resistivity acquires exponential dependence on temperature
\cite{Mott1990,Efros1985}:

\begin{equation}
\rho(T)=\rho_{o}\exp[-(\frac{T_{0}}{T})^{\zeta}]\label{eq:rho_VRH}
\end{equation}
Here the prefactor $\rho_{0}$ is determined by the electron-phonon
matrix element, $\xi$ is the localization radius.

\begin{figure}[ptb]
\includegraphics[width=0.4\textwidth]{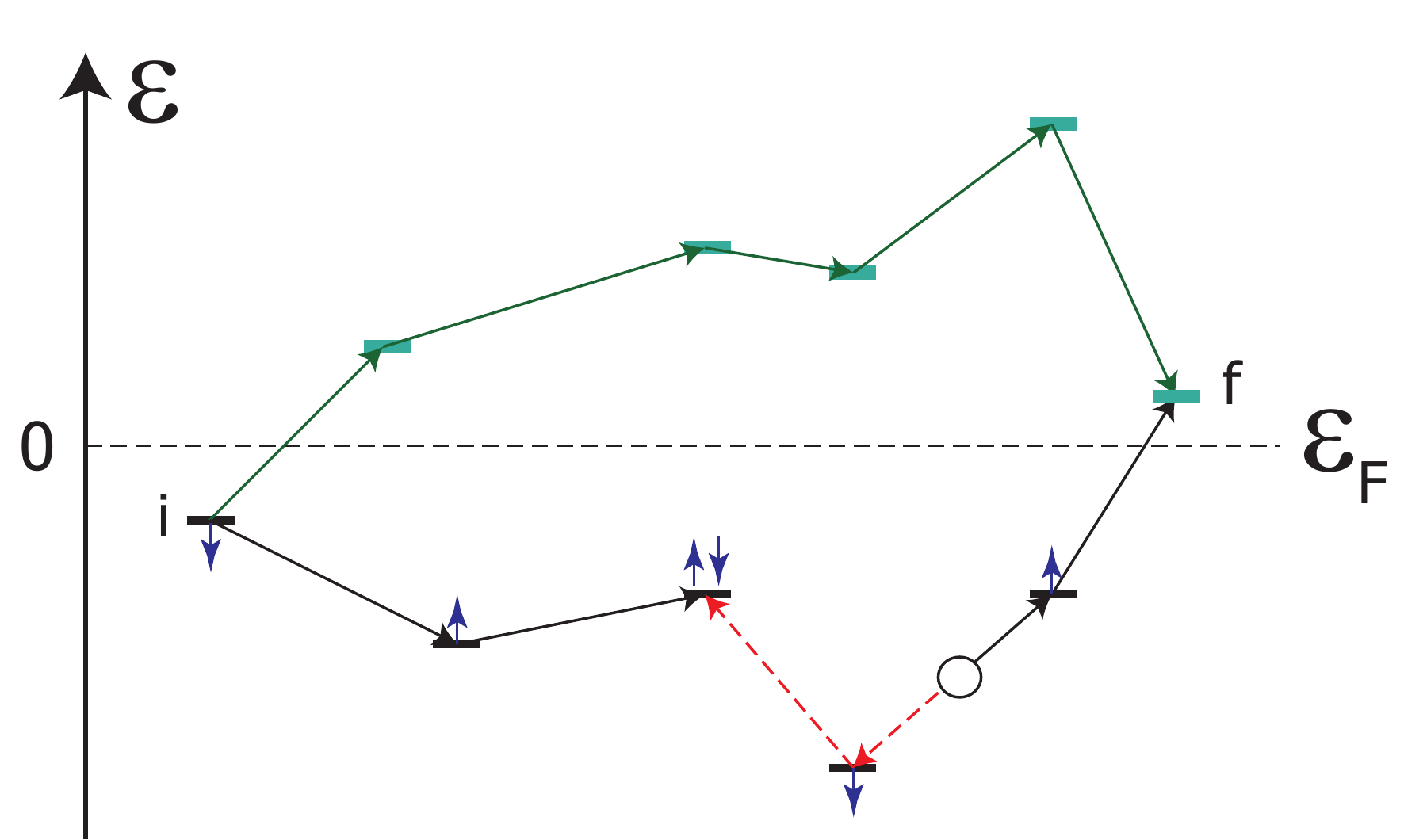} \caption{Qualitative picture of the phonon assisted tunneling through localized
states from the initial state $i$ to the final state $f$. Solid
bars indicate the energies of the localized states. The energies of
the initial and final states are close to the Fermi energy $\epsilon_{F}=0$
(indicated by the dashed line) whilst the intermediate localized states
are typically further away from $\epsilon=0$. The states with negative
energies can be filled with one or two electrons. In the former case
they are characterized by the spin of the electron shown by vertical
arrows. The states with $\epsilon>0$ are empty. Black and green (gray)
arrows indicate electron tunneling paths through empty and filled
localized states. If the path goes through the site that is already
occupied by the electron with the same spin, the coherent process
happens by creating and electron hole pair (indicated by empty circle),
then by tunneling hole carrying the opposite spin in the opposite
direction and finally by anihilating it with the electron coming from
the left. This process leaves the spin state intact. The incoherent
process in which the hole carrying the same spin might be also possible
in some physical situations (see section \ref{sub:Beyond-single-particle}). }

\label{fig:HoppingViaLocalizedStates} 
\end{figure}

Generally, the density of localized states can be energy dependent
close to the Fermi energy \cite{Efros1985}: 

\begin{equation}
\nu(\epsilon)=C\epsilon^{\beta}.\label{eq:densStates}
\end{equation}
where we count the energy, $\epsilon$, of a tunneling electron from
the Fermi energy. In the absence of electron-electron interaction
(Mott's theory) the density of states at the Fermi level is constant$\left(\beta=0,\, C=\nu_{0}\right)$
leading to activation energy $T_{0}\approx13(\nu_{0}\xi^{2})^{-1}$
and exponent $\zeta=1/3$ for $d=2$ (Mott law). In the case when
electrons (in 2D or 3D) interact via three dimensional Coulomb interaction
(Efros-Shklovskii regime) $\beta=1$, $C\approx\left(2/\pi\right)e^{4}/\kappa^{2}$
, where $\kappa$ is the dielectric constant. This results in $\zeta=1/2$,
$T_{0}\sim e^{2}/\kappa\xi$ for 2D electrons interacting via three
dimensional Coulomb. 

The qualitative arguments of the Mott theory can be made more quantitative
by considering the optimal percolating cluster of electron hops.\cite{Efros1985}
Probability of a single hop between the states localized around positions
$r_{i}$ and $r_{j}$ is given by 
\begin{equation}
W_{ij}=\frac{2\pi}{h}\int|M_{ij}\left(\vec{q}\right)|^{2}\delta\left(\epsilon_{i}-\epsilon_{j}-uq\right)d^{d}d\label{eq:W_ij}
\end{equation}
 Here 
\begin{equation}
M_{ij}\sim\int d^{d}r\psi_{i}\left(\vec{r}-\overrightarrow{r}_{i}\right)\psi_{j}\left(\vec{r}-\overrightarrow{r}_{j}\right)e^{i\vec{q}\vec{r}}\label{eq:M_ij}
\end{equation}
is the phonon matrix element, $u$ is the speed of the sound and $\overrightarrow{q}$
is its wave vector. Because the wave functions $\psi_{i}(\vec{r}-\overrightarrow{r}_{i})$
and $\psi_{j}(\vec{r}-\overrightarrow{r}_{j})$ decrease exponentially,
$M_{ij}$ and $W_{ij}$ are exponential functions of the localization
length, $\xi\left(B\right)$.

In the main part of our paper we consider the range of magnetic field
in which $W_{ij}(B)$ dependence is dominated by $\xi\left(B\right)$.
In this case one can approximate the phonon tunneling matrix element
by the amplitude of tunneling between states $i$ and $j$: $M_{ij}\sim A_{ij}$.

In the uniform medium the magnetic field suppresses the amplitude
of a single quantum tunneling event: 
\begin{equation}
A_{ij}\propto\exp(-r_{ij}^{2}/2L_{B}^{2})\mathrm{\, at\,}r_{ij}\gg L_{B}^{2}/\xi
\end{equation}
which gives positive magnetoresistance. Here $L_{B}=\left(c\hbar/eB\right)^{1/2}$is
the magnetic length. In disordered media, electrons scatter from other
localized states which have energies different from the energy of
the final state. The effect of magnetic field is due to the interference
of the directed optimal paths, which is shown schematically in Fig.~\ref{fig:Interference}b.
In this case $A_{if}=\sum_{\Gamma}A_{\Gamma}$ is a coherent sum of
amplitudes, $A_{\Gamma}(B)$, to tunnel along paths $\Gamma$, between
the initial \textquotedbl{}i\textquotedbl{} and final \textquotedbl{}f\textquotedbl{}
sites. The tunneling paths can be defined by the sequence of states
which scatter electrons in the course of tunneling. At zero magnetic
field $\textbf{B}=0$ the wave functions of localized states and the
tunneling amplitudes $A_{\Gamma}(0)$ can be chosen to be real: \cite{Shklovskii1985}

\begin{eqnarray}
A_{if}(0) & = & \frac{1}{|\textbf{r}_{f}-\textbf{r}_{i}|^{1/2}}\exp(-\frac{|\textbf{r}_{j}-\textbf{r}_{i}|}{\xi})+\sum_{\alpha}\frac{1}{|\textbf{r}_{\alpha}-\textbf{r}_{i}|^{1/2}}\exp(-\frac{|\textbf{r}_{\alpha}-\textbf{r}_{i}|}{\xi})\frac{\left(\mu_{\alpha}\right)^{1/2}}{|\textbf{r}_{\alpha}-\textbf{r}_{j}|^{1/2}}\exp(-\frac{|\textbf{r}_{\alpha}-\textbf{r}_{f}|}{\xi})\label{eq:A_ij(0)}\\
 & + & \sum_{\alpha,\beta}\frac{1}{|\textbf{r}_{\alpha}-\textbf{r}_{i}|^{1/2}}\exp(-\frac{|\textbf{r}_{\alpha}-\textbf{r}_{i}|}{\xi})\frac{\left(\mu_{\alpha}\right)^{1/2}}{|\textbf{r}_{\beta}-r_{\alpha}|^{1/2}}\exp(-\frac{|\textbf{r}_{\beta}-\textbf{r}_{\alpha}|}{\xi})\frac{\left(\mu_{\beta,}\right)^{1/2}}{|\textbf{r}_{\beta}-\textbf{r}_{f}|^{1/2}}\exp(-\frac{|\textbf{r}_{\beta}-\textbf{r}_{f}|}{\xi})+....\nonumber \\
 & = & \sum_{\Gamma}A_{\Gamma}(0)
\end{eqnarray}
\begin{equation}
\mu_{\alpha}\sim\frac{b}{\epsilon_{\alpha}-\epsilon_{i}}\label{eq:mu_alpha}
\end{equation}
 Here $\mu{}_{\alpha}$ is the amplitude of scattering on $\alpha$'s
localized state, $\epsilon_{i}$ and $\epsilon_{\alpha}$ are energies
of the tunneling electron and the localized scattering state, $b\sim\sqrt{\xi}\epsilon_{0}>0$,
and $\epsilon_{0}$ is the characteristic binding energy of the localized
states. Generally $\epsilon_{\alpha}$ are random quantities, so the
amplitudes $A_{\Gamma}(\textbf{B}=0)=A_{\Gamma}(0)$ have random signs.
Note that the equation (\ref{eq:A_ij(0)}) describes both the processes
in which an electron is scattered by empty sites and the ones in which
it goes through the occupied sites (see Fig. \ref{fig:HoppingViaLocalizedStates})
which can be described as a hole moving backwards. The important condition
for the interference is that in the final state all intermediate electrons
should return to their original positions and spin states. 

The hopping probability $W_{if}$ is a random quantity. Generally,
to get the value of the resistance of the system one has to solve
the full percolation problem with the probability of individual hops
given by $W_{if}$.\cite{Efros1985} However, as long as $\ln[\rho(B)/\rho(0)]/\ln\rho(0)\ll1$
the magnetoresistance is given by the average of the logarithm of
the hopping probability\cite{Efros1985}: 

\begin{equation}
\ln\frac{\rho(B)}{\rho(0)}=-\left\langle \ln\frac{W_{if}(B)}{W_{if}(0)}\right\rangle \label{eq:ln(rho(B))}
\end{equation}
Here the brackets denote averaging over random scattering configurations
and over different hoppings which belong to a percolation cluster.
These hoppings are characterized by a typical hopping length, $r_{hop}$.
With a good accuracy, one can replace the full average (\ref{eq:ln(rho(B))})
with the average over random scattering configurations for the hopping
processes by the distance $r_{hop}.$ Physically the averaging of
the logarithm in (\ref{eq:ln(rho(B))}) means that the resistivity
is controlled by the typical hopping probability, rather than by rare
events. 

The application of a magnetic field $\textbf{B}$ introduces random
phases to the tunneling amplitudes 
\begin{equation}
A_{\Gamma}(B)=A_{\Gamma}(0)\exp[i2\pi\frac{\Phi_{\Gamma}}{\Phi_{0}}],\label{eq:A_Gamma(B)}
\end{equation}
where $\Phi_{\Gamma}=BS_{\Gamma}$, $S_{\Gamma}$ is the area enclosed
between the path $\Gamma$ and straight line going from initial to
final states, see Fig.~\ref{fig:Interference}b. 

Depending on distributions of the signs of the amplitudes $A_{\Gamma}(0)$
the orbital magnetoresistance can be both positive and negative. To
illustrate this fact let us consider a model in which there are only
two paths, $A_{1}(0)\sim A_{2}(0)$, which are independent random
quantities and $|\Phi_{1}-\Phi_{2}|\sim\Phi_{0}$. If $A_{1,2}(0)>0$
are positive, in the presence of magnetic field, the amplitudes $A_{\Gamma}(B)$
partially cancel each other. As a result, $\langle\ln W_{ij}(B)\rangle$
decreases by a factor of the order of one when $|\Phi_{1}-\Phi_{2}|\sim\Phi_{0}$.
In this case the magnetoresistance is positive. 

The situation changes if $A_{1,2}(0)$ have random signs. In the simplest
case when the signs are completely random, the \emph{average} probability
$\left\langle \left|\sum A_{\Gamma}(B)\right|{}^{2}\right\rangle =\sum\left\langle \left|A(0)\right|^{2}\right\rangle $
is independent of $\textbf{B}$. If magnetic flux through the closed
loop formed by paths $1$ and $2$ is larger than the flux quantum,
the phases of the amplitudes $A_{1,2}$ are completely random, so
that $\left\langle A_{1}(B)A_{2}(B)\right\rangle =0.$ This implies
that the variance $\left\langle \left|\sum_{\Gamma}A_{\Gamma}(B)\right|{}^{4}\right\rangle -\left\langle \left|\sum_{\Gamma}A_{\Gamma}(B)\right|^{2}\right\rangle ^{2}$
decreases by a factor of the order of one when $|\Phi_{1}-\Phi_{2}|\sim\Phi_{0}$.
As a result, a typical value of $W_{if}$ defined by (\ref{eq:ln(rho(B))})
increases by a factor of the order of one and the magnetoresistance
is negative.

This simplified picture of magnetoresistance being determined by the
interference between two paths becomes more complicated for two reasons.
First, at large scales the propagation amplitude is dominated by many
paths which go through the same scatterer or a group of scatterers.
This implies strong correlations between amplitudes $A_{\Gamma}$
, as we discuss in section \ref{sub:Predictions-of-single}. This
makes the mathematical problem of calculation of $\rho(B)$ non-trivial.
Second, the behavior of the magnetoresistance becomes more complicated
if amplitude signs are correlated at some finite distances (see section
\ref{sub:The-sign-transition.}). In this case, one expects a crossover
from the negative to positive magnetoresistance as the field is increased,
as we explain in section \ref{sec:Magnetoresistance-in-the}. 

Because the sign and the magnitude of the magnetoresistance are intimately
related to the statistics of sign and amplitude distribution of $A_{ij}(0)$
we start with a discussion of this quantity.

\subsection{Statistics of the amplitude $A$ in the absence of the magnetic field.\label{sub:Statistics-of-the}}

In the case of small and positive scattering amplitudes, $\mu_{\alpha}>0$,
and at zero magnetic field the problem of electron tunneling can be
mapped \cite{Medina1992,Medina1996,Kim2011,Shklovskii1985} onto the
problem of directed polymers. In the latter problem one studies the
thermodynamics of an elastic string in a delta-correlated two dimensional
random potential, $W(x,y)$ that is characterized by energy functional
\begin{equation}
H_{dirpol}\{y(x)\}=\int_{-\infty}^{x}\left[\frac{\sigma}{2}(\partial_{x}y)^{2}+W(x,y(x))\right]dx\label{eq:H_pol}
\end{equation}
Introducing the partition function, $Z(y,x)=\sum_{y\{x\}}\exp(-\beta H)$
of the string that ends at point $(x,y)$ one gets that its evolution
as a function of $x$ is described by the equation
\begin{equation}
\partial_{x}Z=\frac{1}{2\beta\sigma}\partial_{y}^{2}Z-\beta W(x,y)Z.\label{eq:d_xZ}
\end{equation}
This equation should be compared with the equation for the particle
propagation in disordered media:
\begin{equation}
E\Psi=-\frac{1}{2m}\nabla^{2}\Psi+V(x,y)\Psi.\label{eq:EPsi}
\end{equation}
with white noise potential $V(x,y)$. At negative energies corresponding
to tunneling we substitute $\Psi=\exp(-\beta\sigma x)Z(x,y)$; one
can neglect second order derivative in $x$ terms that are small at
weak potential $V\ll-E$. Then, the Schrodinger equation (\ref{eq:EPsi})
coincides with (\ref{eq:d_xZ}) with $(\sigma\beta)^{2}=-2mE$ and
$W=\sigma V/2E$. This mapping also holds for arbitrary (not necessarily
white noise correlated) potential $V$. However, it becomes less useful
for arbitrary potentials because analytical results for this problem
were obtained only in the case of the white noise potential. 

Computation of positive magnetoresistance requires the solution of
the directed polymers beyond the white noise approximation, so the
analytical results are not directly applicable. Furthermore, the physically
relevant problem of scattering with negative amplitudes cannot be
mapped onto any thermodynamic problem because the corresponding free
energy becomes imaginary. The applicability of the results of the
directed polymer problem in the white noise approximation becomes
even more questionable in this case. Below we give a brief review
of the results of the direct polymers problem in the white noise approximation.
Then we present results of our numerical simulations beyond the white
noise approximation, which indicate that these problems belong to
the same universality class. Finally, we discuss the statistics of
the signs of the tunneling amplitude and show that the existence of
the ``sign phase transition'' is compatible with the results for
directed polymer problem.

The main result of the directed polymer theory is the scaling form
of the fluctuational part of the free energy of the polymer of length
$L$, $F\propto L^{1/3}$and its deviations in the transverse direction
$Y\propto L^{2/3}.$ For equivalent problem of domain wall pinning
this scaling was first found numerically in work\cite{Huse1985}.
Analytically, it was extracted from the third moment of the distribution
function of polymers of length $L$, ${\cal {P}}(F)$.\cite{Kardar1987a,Kardar1987b}
The replica method that was used in this work might be questioned
because of an apparent non-commutativity of the limits $L\rightarrow\infty$
and $n\rightarrow0$ and because it gives unphysical results for all
moments of the distribution function except the third. All these problems
can be eliminated by solving for the distribution of the energy differences
of the infinitely long polymers that end at different points $y_{1},y_{2}$,
this solution gives the same scaling exponents. \cite{Dotsenko2008}
as the original approach\cite{Huse1985,Kardar1985a,Kardar1985b,Kardar1986,Kardar1987a,Kardar1987b}. 

The striking generality of this scaling result that we prove by numerical
simulations below is, probably, due to the qualitative reasoning that
relates it to the Markovian form of the free energy fluctuations as
a function of transverse coordinate. Indeed, The Markovian form implies
that free energy fluctuations at large scales are proportional to
$Y^{1/2},$ on the other hand they should be of the order of the string
elastic energy at these scales: $Y^{2}/L\propto Y^{1/2}$. Solving
the last equation for $Y$ we get the scaling dependencies of the
exact solution and of the numerical simulations. 

Despite being intuitively appealing, the Markovian nature of free
energy fluctuations is difficult to prove for the physically relevant
situation in which some scattering amplitudes (\ref{eq:mu_alpha})
are very large. It is even more difficult to prove it for the case
of rare negative scattering amplitudes in which wave function can
change sign at some points. At these points the free energy defined
by $F\equiv-T\ln Z$ acquires imaginary part ($\Im F=\pi$) whilst
its real part becomes large. Because these points are due to close
by negative scatterers, the effective free energy becomes highly correlated
which violates the main assumption of the Markovian nature of the
free energy fluctuations.

Recently\cite{Calabrese2010,Dotsenko2010a}, a full Bethe ansatz solution
of problem (\ref{eq:H_pol}) established the complete form of the
distribution function of free energy $F\equiv-T\ln Z$ of the string
of length $L$, which turns out to coincide with the Tracy-Widom distribution\cite{Tracy1994}.
This result allows one to check if the problem of particle hopping
belongs to the same universality class as the directed polymers. Namely,
we define the effective free energy of the quantum problem as 
\begin{equation}
F=-\Re\ln A(x,y)\label{eq:ReF}
\end{equation}
where $A$ is the electron amplitude at site $(x,y)$ propagating
in $x$-direction. This free energy describes the decay of the wave
function. We compute the amplitude $A$ by simulating electron propagation
and check the scaling properties of its real part fluctuations in
$y$-direction and the universality of the distribution function. 

We determine the amplitude $A$ from the solution of the lattice recursive
equation

\begin{equation}
A_{i,j}=\frac{g}{\epsilon_{ij}}[A_{i-1,j+1}+A_{i-1,j}+A_{i-1,j-1}]\label{eq:A_ij}
\end{equation}
where $\epsilon_{ij}$ are random independent variables defined on
each lattice site and $g$ is the parameter that determines the average
decay of the amplitude (inverse localization length). We shall discuss
below different distribution functions of $\epsilon_{ij}$ appropriate
for different physical systems. 

\begin{figure}[ptb]
\includegraphics[width=0.4\textwidth]{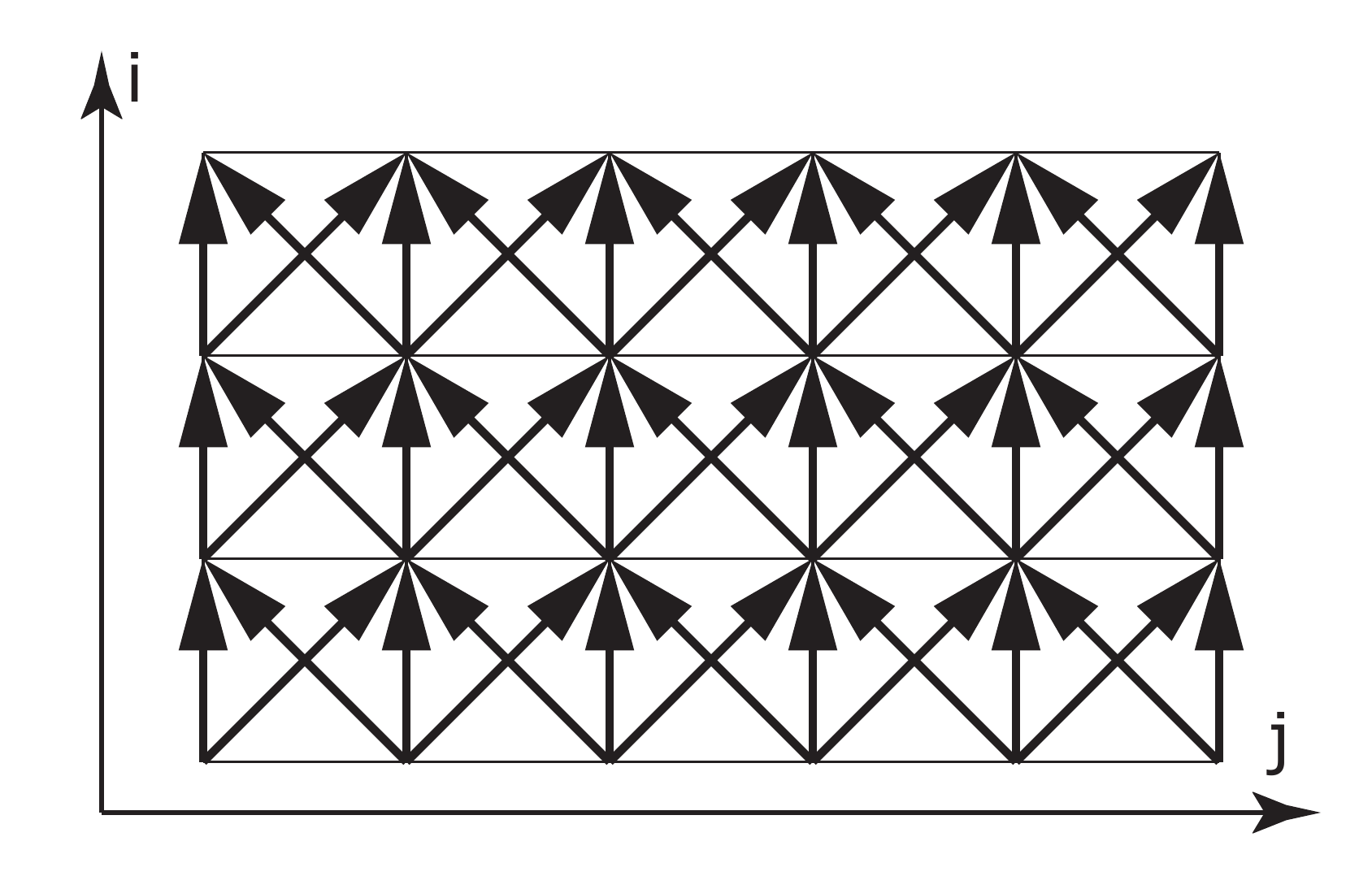} \caption{Schematics of the electron propagation described by equation (\ref{eq:A_ij}).
The computation of the localization length discussed in section \ref{sub:Statistics-of-the}
involved simultaneous propagation of amplitudes in the vertical direction
for many (typically $L>10^{6}$) steps. For the computation of the
matrix elements in section \ref{sub:Prediction-for-magnetoresistance}
the wave functions were assumed to be localized on two sites in the
middle of upper and lower rows at distance $L$ and then determined
in the middle . }

\label{fig:NumericalScheme} 
\end{figure}

Physically, the model (\ref{eq:A_ij}) describes the motion of electrons
on the lattice shown in Fig. \ref{fig:NumericalScheme}. The site
with energy $\epsilon_{ij}=\left\langle \epsilon\right\rangle $ can
be identified with ideal lattice, the rest with impurities. If energy
$\epsilon_{ij}$ is distributed in a narrow interval around its average,
the evolution (\ref{eq:A_ij}) becomes equivalent to (\ref{eq:EPsi})
in the continium limit. As discussed in section \ref{sub:Review-of-variable}
the most physically natural choices of the distribution function of
$\epsilon$ are uniform $P(\epsilon)=\theta(\epsilon)$, linear $P(\epsilon)=2\epsilon$
and their analogs for the negative scattering amplitudes: $P(\epsilon)=1/2$,
linear $P(\epsilon)=|\epsilon|$. In all cases we assume that the
distribution is cutoff by $\epsilon_{0}$ at large $\epsilon$: $P(|\epsilon|>\epsilon_{0})=0$.
The choice of $\epsilon_{0}$ determines the average decay rate of
the electron amplitude which is mostly irrelevant, in the computations
we have set it to $\epsilon_{0}=1$. We have also studied the gapped
distribution $P(\epsilon)=2$ for $1/2<\epsilon<1$ for which we expect
to get the results similar to the one predicted by exact solution.
Finally we studied the binary distribution $P(\epsilon)=(1-X)\delta(\epsilon-1)+X\delta(\epsilon-(\mu+1)^{-1})$
characterized by parameter $X$ and negative scattering amplitude
$\mu<0$.

\begin{figure}[ptb]
\includegraphics[width=0.4\textwidth]{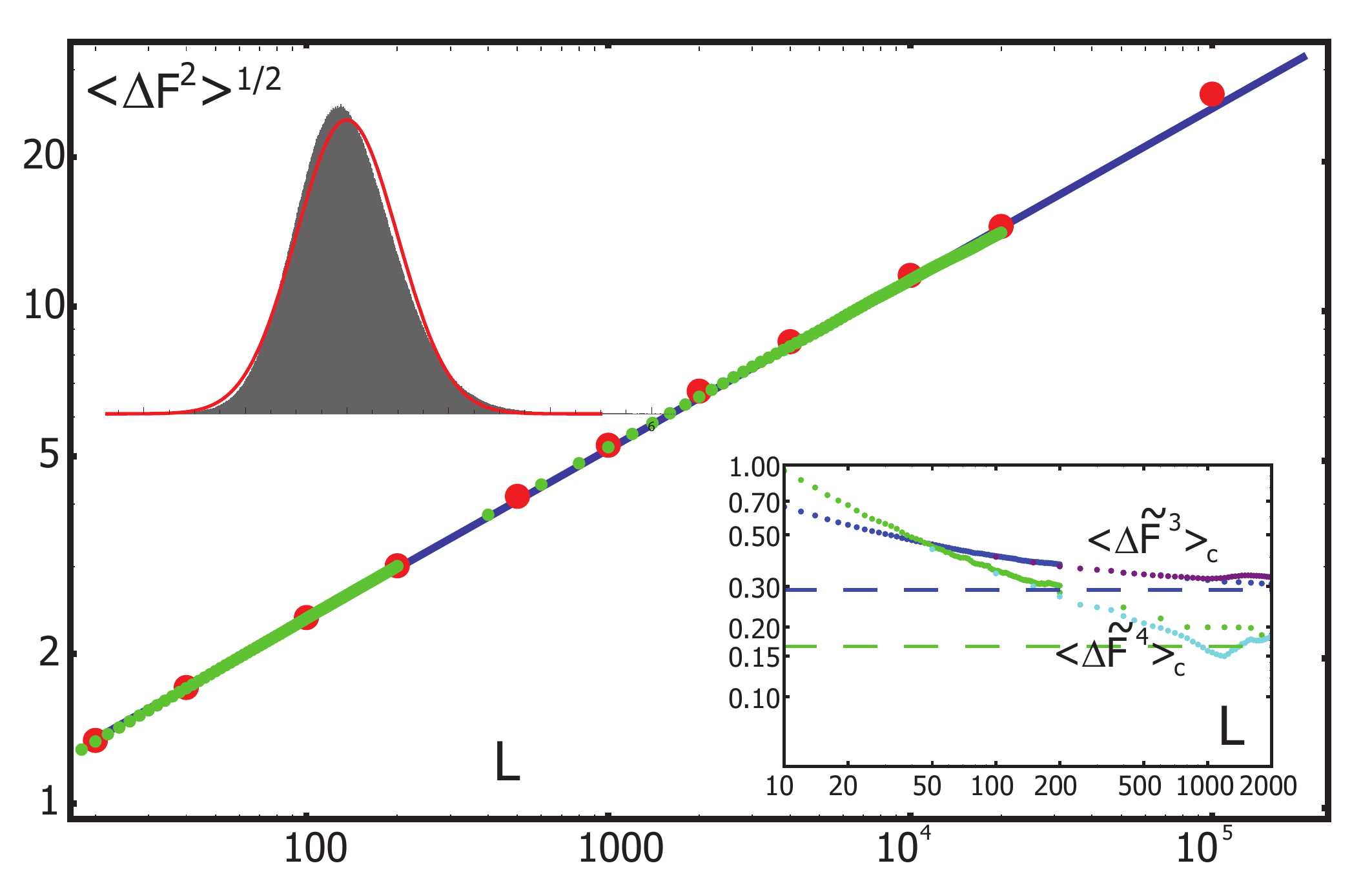}\includegraphics[width=0.4\textwidth]{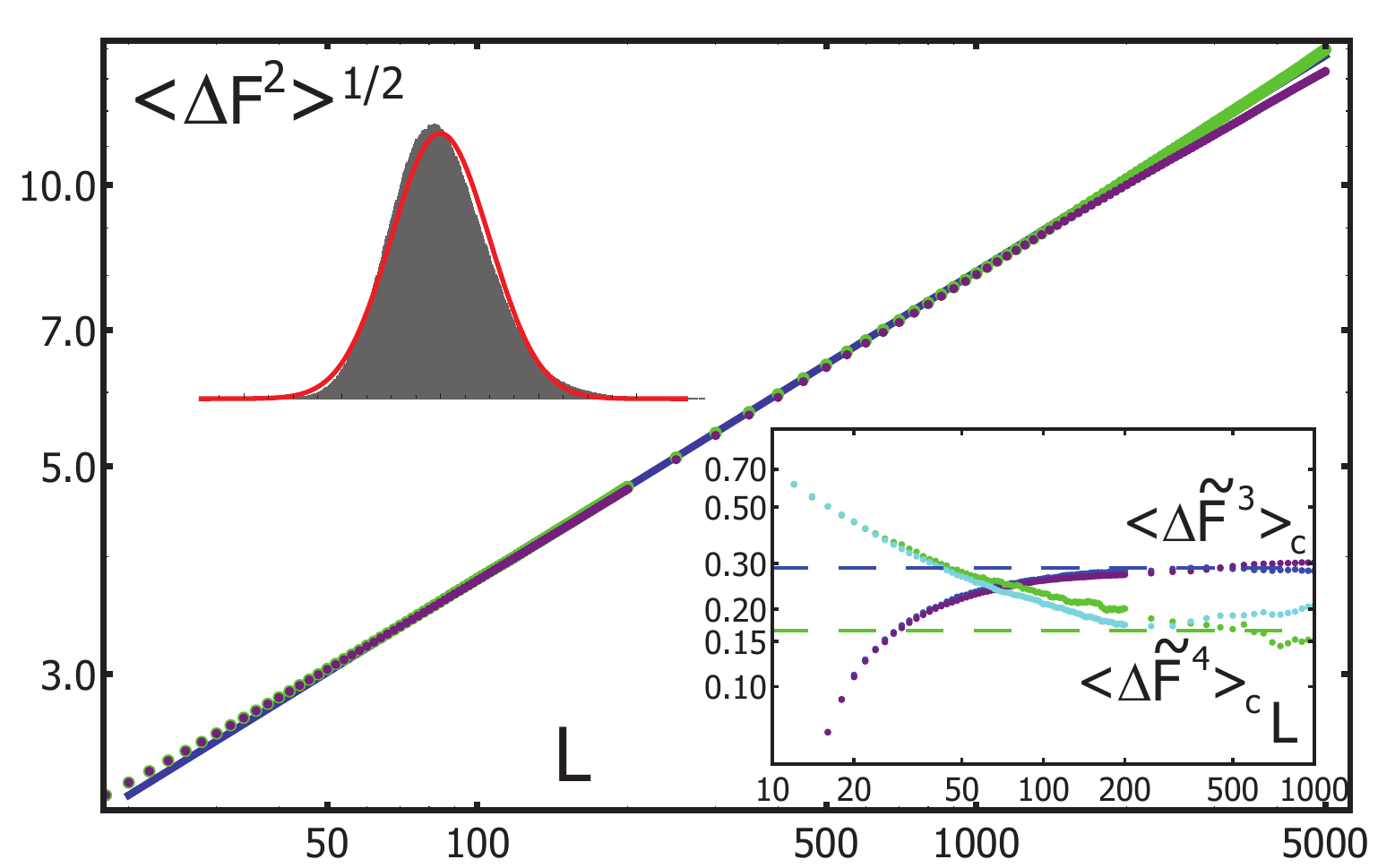}
\caption{Scaling dependence of the fluctuations of electron wave function decay,
$\Delta F=\left\langle F\right\rangle -F$ where $F$ is defined by
(\ref{eq:ReF}). The quantity $F$ is equivalent to the free energy
of the directed polymer problem. The left panel shows the results
for the linear density of states with $P(\epsilon<0)=0$, the right
panel gives the same results for the equally probable positive and
negative scattering amplitudes. The upper inserts show the distribution
function of $\Delta F$ and its fit to the Gaussian compared to which
the distribution is slightly skewed as expected for Tracy-Widom distribution.
The lower inserts show the evolution of the normalized higher moments
of the distribution function that tends to the universal values expected
for Tracy-Widom (shown as dashed horizontal lines). The numerical
results were obtained by by simulating the evolution (\ref{eq:A_ij})
on systems of sizes $N=10^{6},10^{7}$ and $5\times10^{7}$as indicated
by points of different size and colors. The straight line correspond
to exponent $\gamma=0.345$ and $0.33$ for left and right panels
respectively. The convergence to the scaling form of the free energy
fluctuations happens relatively fast while higher moments of the distribution
function require enormous statistics, especially at large $L$ as
is indicated by the deviation of curves representing fourth moment
for $N=10^{7}$ and $N=5\times10^{7}$}

\label{fig:Delta kappa distribution for linear DOS positive signs} 
\end{figure}

\begin{figure}[ptb]
\includegraphics[width=0.4\textwidth]{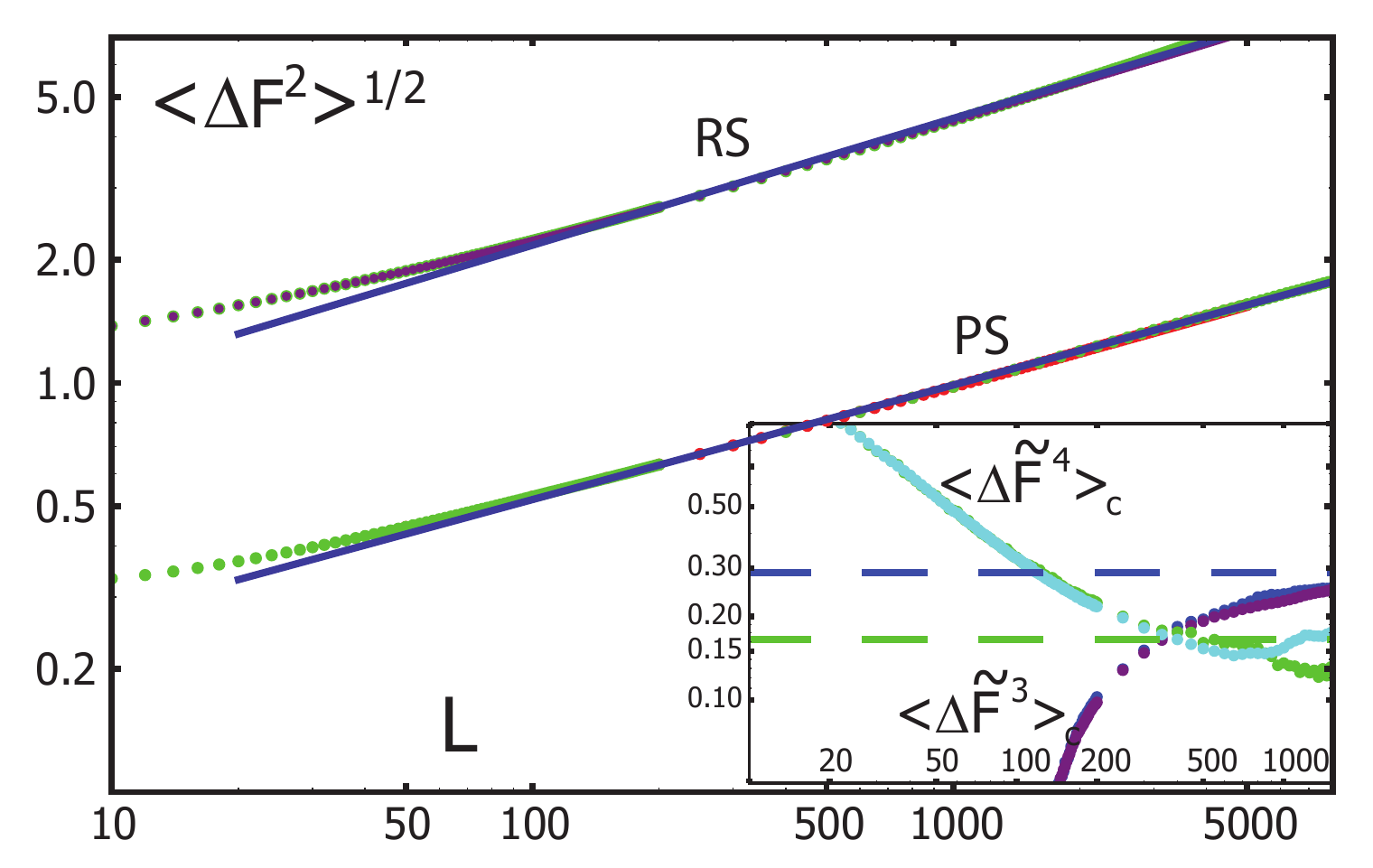}
\caption{Scaling dependence of the fluctuations of electron wave function decay,
$\Delta F=\left\langle F\right\rangle -F$ obtained from the numerical
solution of the evolution (\ref{eq:A_ij}) with gapped density of
states. The lower data set (denoted PS) correspond to the positive
scattering amplitudes, the upper data set (denoted RS) to the completely
random amplitudes with equal probability of signs. The data were fit
with the scaling dependencies with the exponent $\gamma=0.28$ for
positive scatterers and $\gamma=0.31$ for random signs. The results
were obtained for the systems of size $N=10^{7}$ and $5\times10^{7}$.
Higher moments tend to the universal values of Tracy-Widon distribution
as shown in the insert that gives the data for random sign scatterers. }

\label{fig:Delta kappa distribution for gapped DOS} 
\end{figure}

Some of our results are presented in Figs.~\ref{fig:Delta kappa distribution for linear DOS positive signs}
and \ref{fig:Delta kappa distribution for gapped DOS}. For all studied
distribution we observe very good scaling, $\left\langle \Delta F^{2}\right\rangle ^{1/2}\propto L^{\gamma}$,
with the exponents $\gamma=0.28,\,0.345,\,0.343$ for gapped, linear
and uniform densities of states respectively. These values are very
close to the expected value $1/3$, especially for the linear and
uniform densities of states. The data for the gapped density of states
display a significant transient regime, so the deviation of the exponent
from the analytical result is not surprising. The presence of negative
scattering amplitudes has small effect on these exponents, they become
$\gamma=0.31,\,0.33,\,0.345$ that are even closer to the expected
values. Furthermore, the higher moments of the distribution function
tend to the universal values expected for the Tracy-Widom distribution.
These results is in agreement with the works\cite{Somoza2007,Prior2009}
that observed Tracy Widom distribution of conductances in two dimensional
models. 

These data lead to the conclusion that the main results of the directed
polymer problem, namely, the scaling dependence of the free energy
and the universality of the distribution function remain valid for
the problem of electron tunneling in disordered media.

\subsection{The sign phase transition. \label{sub:The-sign-transition.}}

As explained in section \ref{sub:Review-of-variable}, the sign of
the magnetoresistance is related to the statistics of signs of amplitudes
$A_{if}(0)$ in the absence of magnetic field. If the concentration
of impurities with negative scattering amplitudes is large, the sign
of $A_{if}(0)$ becomes completely random. If all impurities are characterized
by positive scattering amplitudes $\mu_{i}>0$, the sign of $A_{if}(0)$
is positive. Let us denote the probability to find a positive amplitude
$A_{if}(0)$ by $P_{+}$ and negative by $P_{-}.$ The quantity $\Delta P=P_{+}-P_{-}$characterizes
the sign order. As the concentration, $X$, of the impurities with
negative scattering amplitudes increases, $\Delta P$ should change
from $1$ to $0$. Generally, $\Delta P$ is scale dependent and acquires
its limiting value at $|r_{i}-r_{f}|\rightarrow\infty$. There are
two logical possibilities: either at large scales $\Delta P_{r\rightarrow\infty}=0$
only for $X>X_{c}$ while for smaller $X<X_{c}$ $\Delta P_{r\rightarrow\infty}>0$
, or that any non-zero $x>0$ leads to $\Delta P_{r\rightarrow\infty}=0$.
The former implies that the change in the $x$-dependence of the sign
statistics can be viewed as a phase transition. This possibility has
been suggested in \cite{Nguen1985,Shklovskii1990a,Shklovskii1991},
the alternative was argued for in works \cite{Medina1992,Medina1996,Kim2011}.

\begin{figure}[ptb]
\includegraphics[width=0.4\textwidth]{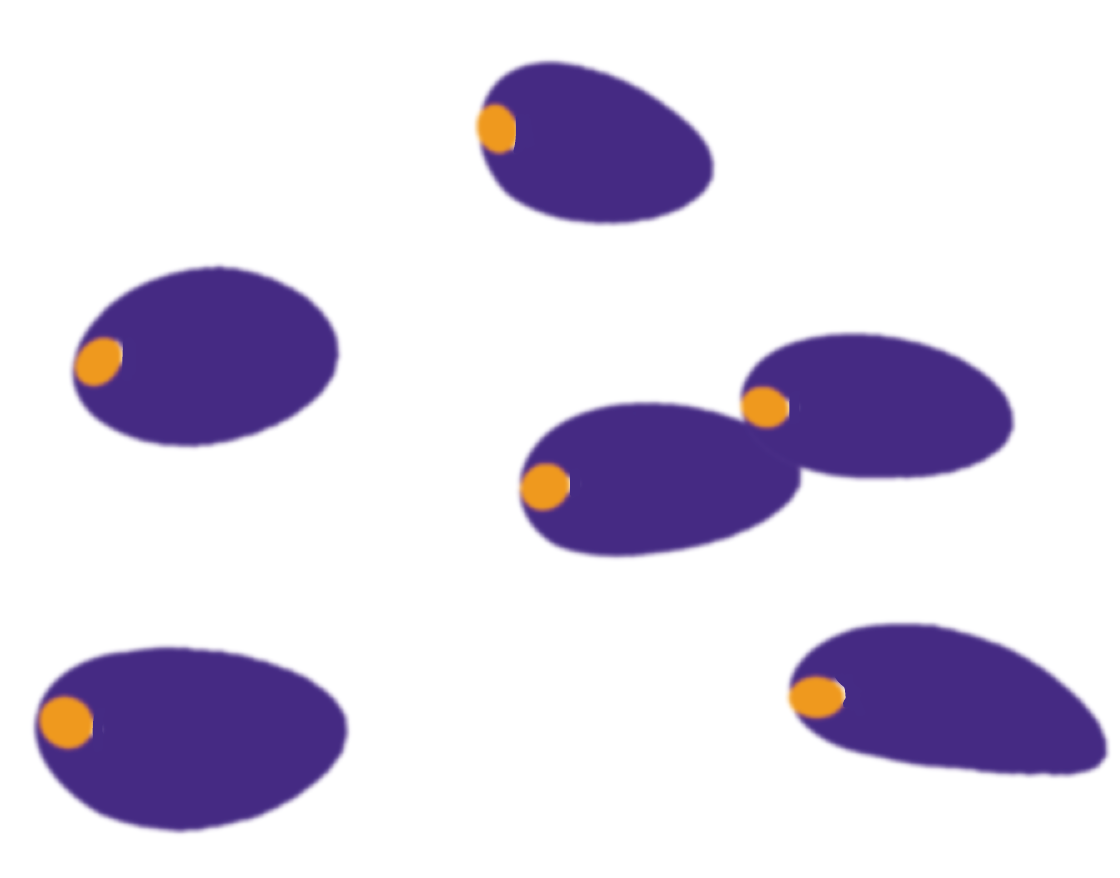} \caption{Qualitative picture of lakes of negative amplitude signs formed in
the wake on an impurity (shown as a small yellow circle) characterized
by negative scattering amplitude}

\label{fig:NegativeLakes} 
\end{figure}

Here we study the sign statistics in the lattice models defined by
(\ref{eq:A_ij}) in section \ref{sub:Statistics-of-the} and show
that both the phase transition and crossover can be realized depending
on the distribution of $\epsilon.$ We start with the simplest case
of binary distribution $P(\epsilon)=(1-X)\delta(\epsilon-1)+X\delta(\epsilon+\epsilon_{0})$
with small $X\ll1$ and small $\epsilon_{0}\ll1$. This model describes
the wave function propagation on the ideal lattice (sites with $\epsilon=1$)
which contains rare impurities characterized by a negative scattering
amplitude $\mu\approx-1/2\epsilon_{0,}|\mu|\gg1$. The large value
of $|\mu|$ allows a continuous description of the tunneling amplitude.
The size of the region where the tunneling amplitude $A_{if}(0)<0$
is negative may be found by noticing that the wave function 

\[
\Psi(x,y)=\exp(-x/\xi)+\frac{\mu}{(x^{2}+y^{2})^{1/4}}\exp(-\sqrt{x^{2}+y^{2}}/\xi)
\]
changes its sign in the egg-shaped region in the wake of the impurity
given by:
\[
y^{2}(x)=x\xi\ln\left[\mu^{2}/x\right],\;0<x<\mu^{2}.
\]
The area of this region is 
\[
S(\mu)=\frac{2}{3}\sqrt{\frac{2\pi}{3}}|\mu|^{3}\xi^{1/2}.
\]
A small concentration, $XS\ll1$ of such impurities leads to independent
lakes of negative signs shown in Fig.~\ref{fig:NegativeLakes}. In
this situation $\Delta P>0$ . 

\begin{figure}[ptb]
\includegraphics[width=0.8\textwidth]{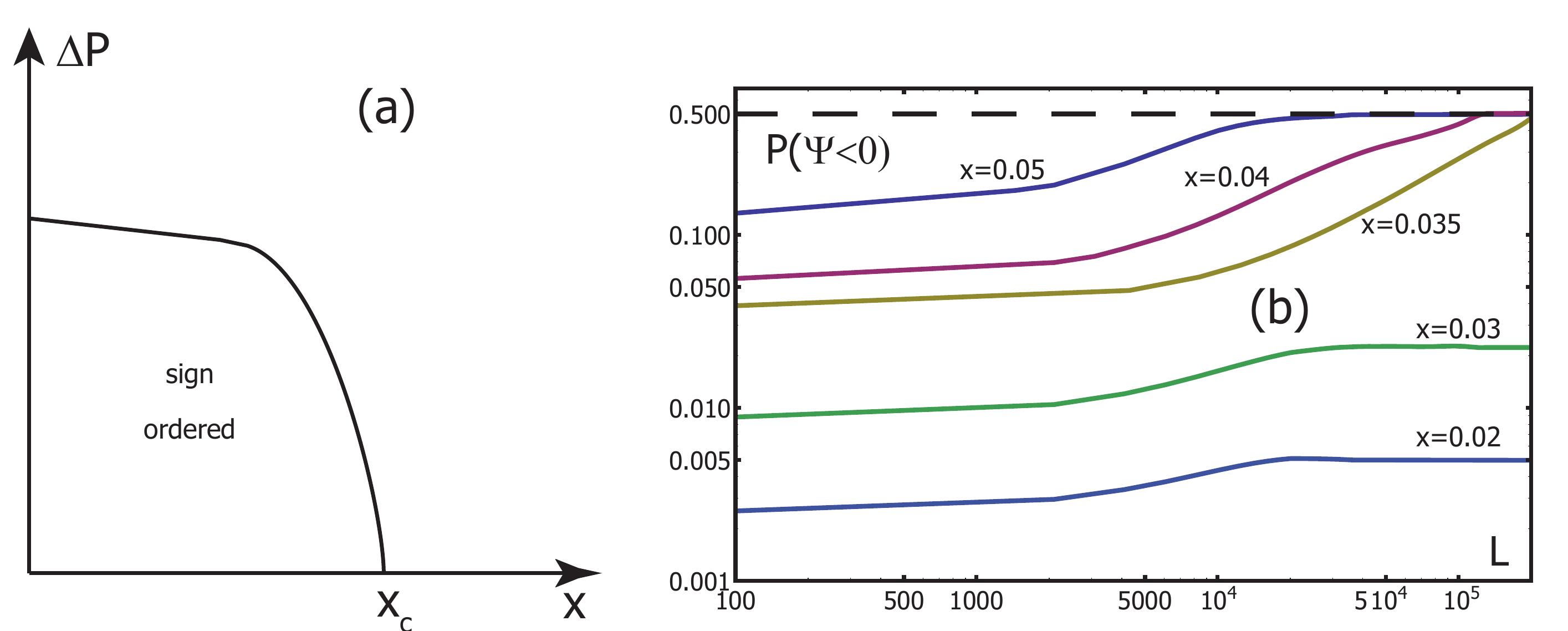} \caption{Panel a: Qualitative picture of the phase transition described by
the order parameter defined $\Delta P(X)$ that happens at $X_{c}\approx0.032$
for the the binary distribution described in the text. Panel b: Scale
dependence of the probability of negative amplitude that shows the
transition around $X_{c}\approx0.032$. }

\label{fig:SignProbability} 
\end{figure}

As the concentration $x$ is increased, different lakes start to overlap
and form a state with random sign of the amplitudes. The transition
between these two phases takes place at $X=X_{c}\sim{\it S}^{-1}\propto|\mu|^{-3}$.
The dependence $P_{-}(X)$ is expected to have a general form characteristic
of a phase transition sketched in Fig.~\ref{fig:SignProbability}a.
These qualitative arguments ignore the contributions from impurities
located close to each other which should not be relevant in the limit
$X\rightarrow0$. 

The numerical simulations show that the transition survives for not
so large values of the scattering amplitudes as well. In particular,
this has been observed for the binary distribution functions with
$\epsilon_{0}=1$. Fig.~\ref{fig:SignProbability} represents the
results of our numerical simulations for this case. As one can see,
the behavior of $\Delta P$ as a function of the distance changes
qualitatively as one increases $X$ ~beyond $X_{c}\approx0.032$.
For smaller concentrations, $x$, probability difference $\Delta P$
saturates at non-zero values, whilst for larger concentrations it
approaches $0$. The scales needed to observe this change in the behavior
are generally very long. We believe that this is the reason that prevented
establishing unambiguously the existence of the transition in early
numerical simulations. We note that the scales are further enlarged
near $X_{c}\approx0.032$ as one expects at a phase transition.

We have also checked that the phase transition between the sign ordered
and sign disordered phases survives for a gapped distribution of $\epsilon$
defined in section \ref{sub:Statistics-of-the}. The numerical data
look very similar to those shown in Fig. \ref{fig:SignProbability},
the expected value of $X_{c}$ in this model is $X_{c}\approx0.02$. 

The existence of the sign phase transition has been questioned in
paper \cite{Kim2011} which used the mapping to the directed polymer
problem. The essence of the argument is that the free energy of directed
polymers leading to a given site are dominated by a single path, so
that just a single impurity along this path suffices to change the
sign of the amplitude. At a small concentration of negative scatterings,
one concludes that the amplitude should become completely random at
the scale $L\propto1/X$. This argument, however, does not take into
account the contribution from subdominant paths that may eventually
restore the sign of the amplitude at large scales as is indicated
by numerical data for the gapped density of states, see section \ref{sub:Statistics-of-the}.

We now show that for a gapless density of states (\ref{eq:densStates})
with $\beta<2$ and for any non-zero concentration of negative scatterers
the sign of the amplitude $A$ becomes completely random at large
scales. Indeed, in this case the total area of negative lakes is

\[
S_{tot}\sim X\int d\epsilon\nu\left(\epsilon\right)S\left[\mu\left(\epsilon\right)\right]
\]
where $S\left[\mu\right]\propto\mu^{3}\propto\varepsilon^{-3}$. Thus,
$S_{tot}$ diverges for all densities of states $\nu\left(\epsilon\right)\sim\varepsilon^{\beta}$
with $\beta\leq2$. This is the case for example in the case of Coulomb
gap where $\nu\left(\epsilon\right)\propto\varepsilon$.

\begin{figure}[ptb]
\includegraphics[height=2in]{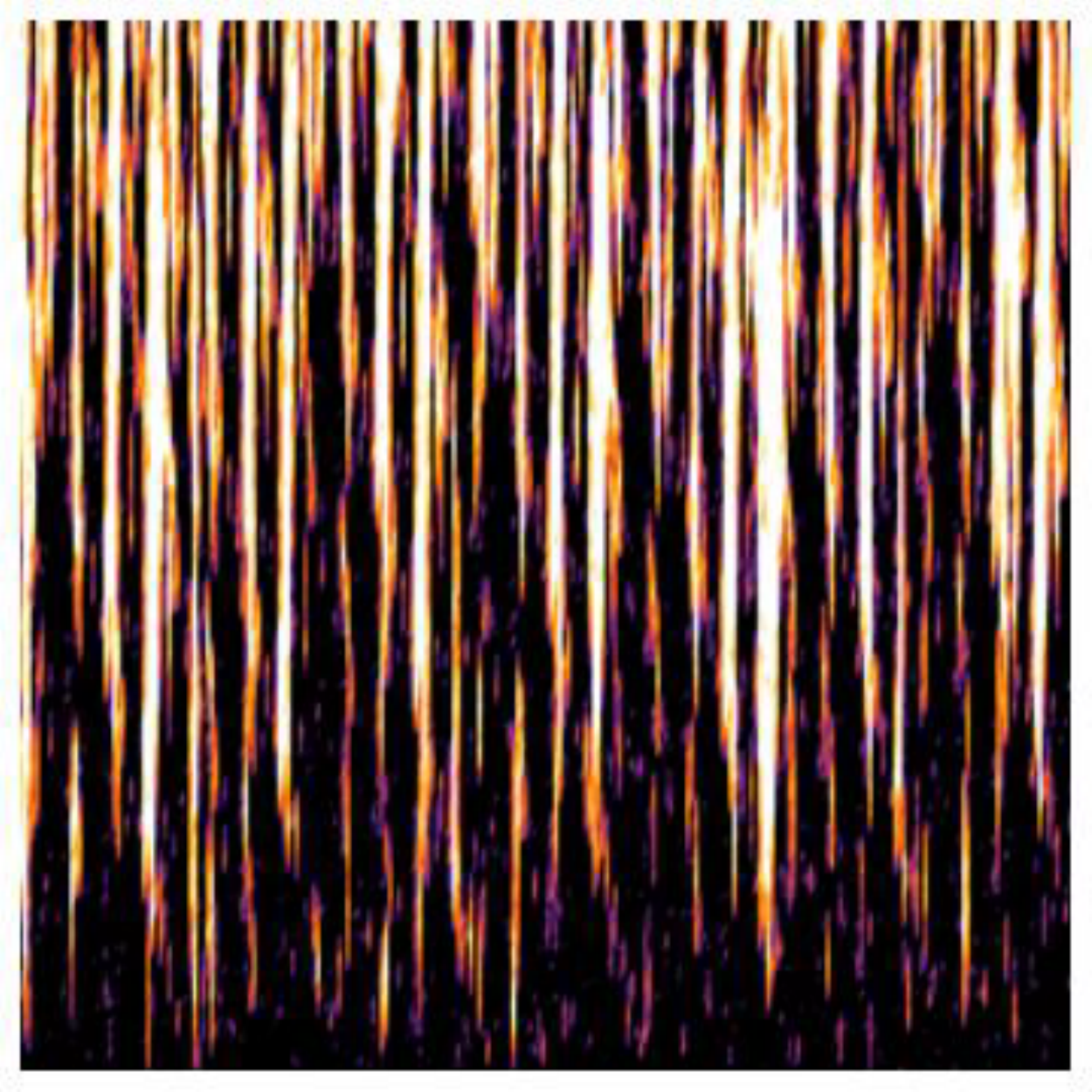}\includegraphics[height=2in]{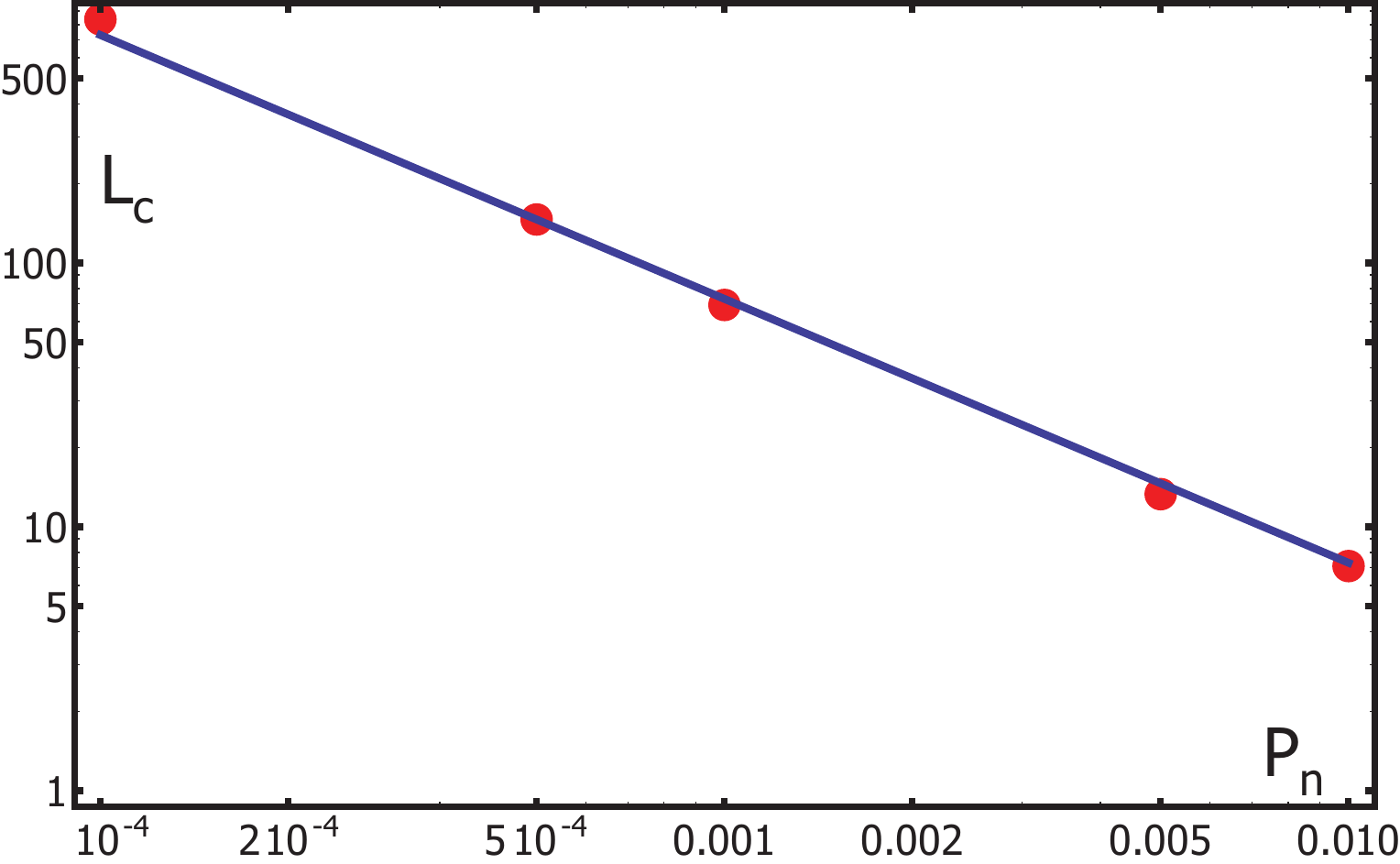}
\caption{The map of the amplitude sign resulting from the wave function evolution
in vertical direction for the linear density of states. The wave function
has all positive signs (shown in black) in the beginning of the evolution
(bottom). As the evolution goes upward the presence of a small concentration
, $X=10^{-4}$, of negative scattering amplitudes results in a larger
and larger regions of negative signs (white regions) until the whole
amplitude sign becomes completely random at the top. The right panel
shows the length scale $L(X)$ at which the sign becomes random as
a function of the concentration $X$. Here we defined $L(X)$ as the
length at which $\Delta P=0.25$. The data are well fit with the dependence
$L\propto1/X$ in agreement with the theoretical expectations based
on directed polymer mapping. }

\label{fig:SignMap} 
\end{figure}

We have checked this conclusion numerically for the linear density
of states and we have indeed observed that even a very small $X\sim10^{-4}$
leads to a random sign of the amplitude at very large scales. Our
data are shown in Fig. \ref{fig:SignMap}. As one expects, the scale
at which the sign becomes random grows quickly with the decrease of
$X$. 

\section{Magnetoresistance in hopping regime.\label{sec:Magnetoresistance-in-the}}

\subsection{Magnetic field dependence of the localization length. \label{sub:Predictions-of-single} }

We now turn to the discussion of magnetoresistance in the variable
hopping regime. We begin by summarizing the results of numerical simulations
for the recursive equation (\ref{eq:A_ij}) that was modified to include
the phases, $\phi_{j}=Bj$, induced by magnetic field 
\begin{equation}
A_{i,j}(B)=\frac{1}{\epsilon_{ij}}[A_{i-1,j-1}e^{i\phi_{j-1/2}}+A_{i-1,j}e^{i\phi_{j-1/2}}+A_{i-1,j+1}e^{i\phi_{j+1/2}}].\label{eq:A_ij(B)}
\end{equation}
Then we give the qualitative explanation of the results based on the
mapping to the directed polymer problem. The dimensionless magnetic
field $B$ in this equation and in the discussion below is given by
the flux of the physical magnetic field, $B_{phys}$, through the
elementary square Malplaquet of the lattice: $B=B_{phys}a^{2}/\Phi_{0}$
where $a$ is the lattice constant and $\Phi_{0}=hc/e$ is the flux
quantum.

Our main result is that at large $r_{hop}>L_{B}$ (which holds at
low temperatures), both positive and the negative magneto resistances
are described by corrections to the localization length: 
\begin{equation}
g(B)=\frac{\Delta\xi(B)}{\xi(0)}=\pm C_{\pm}\left(\frac{B\xi^{2}}{\Phi_{0}}\right)^{\alpha}.\label{eq:UniversalCor}
\end{equation}
This scaling law is characterized by the universal exponent $\alpha\approx4/5$
and non-universal numerical coefficients $C_{\pm}$. The latter depends
on the distribution of $\varepsilon_{ij}$, e.g. $C_{+}\approx2.6$
for the gaped and $C_{+}\approx0.9$ for linear density of states.
Here we define the localization length as the limiting behavior of
the amplitude $\xi=\lim_{r_{ij}\rightarrow\infty}\ln A_{ij}(B)/r_{ij}$.
The positive sign ($+$) in (\ref{eq:UniversalCor}) corresponds to
the case where the system is in the sign disordered phase, the negative
sign corresponds to the sign ordered phase. The universal regime (\ref{eq:UniversalCor})
is achieved at low fields. Notice that whilst the value of $\xi(B)$
is mathematically defined for any magnetic field, its applicability
to the hopping problem requires that $r_{hop}>l_{B}$. 

\begin{figure}[ptb]
\includegraphics[height=3in]{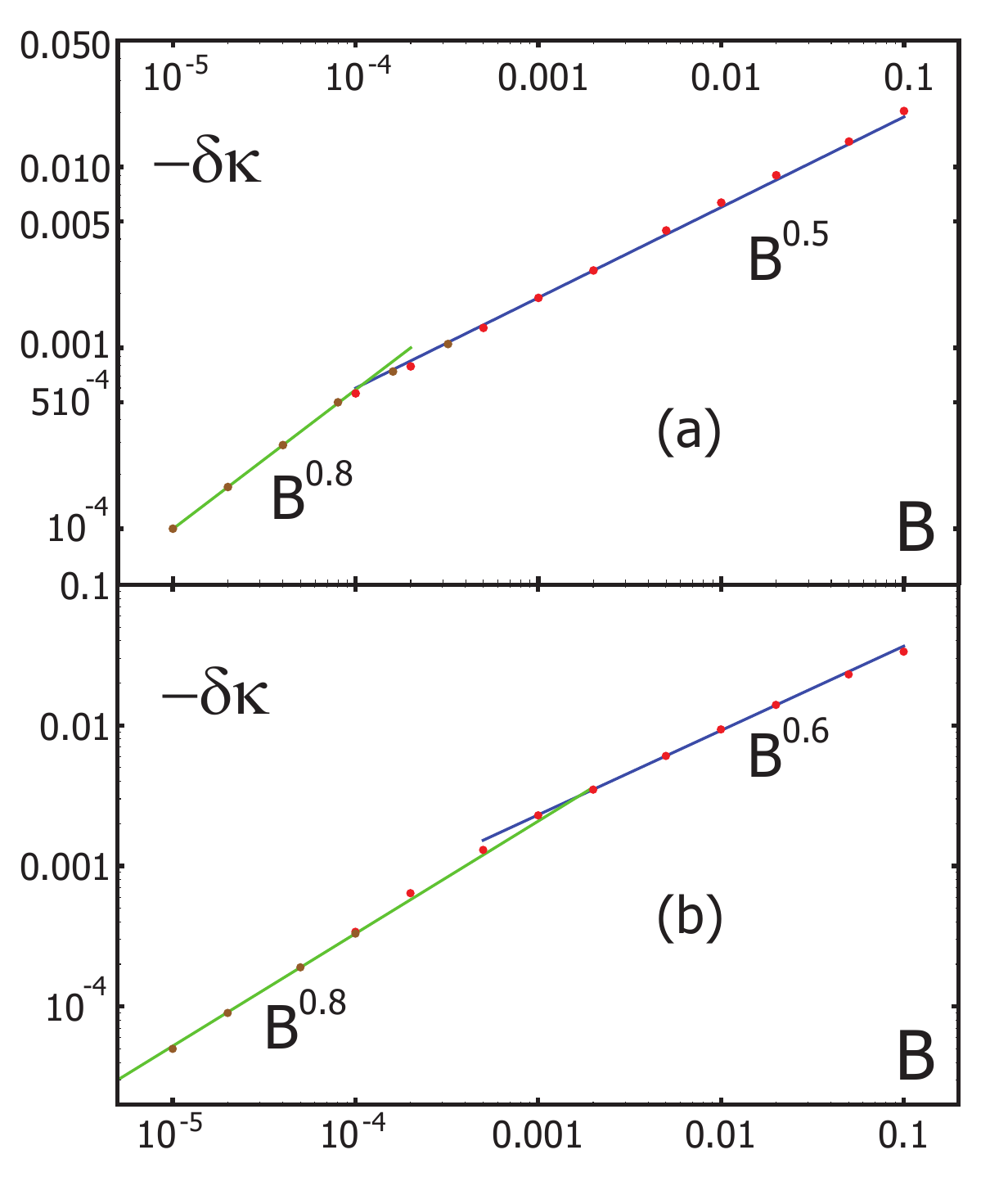}\includegraphics[height=3in]{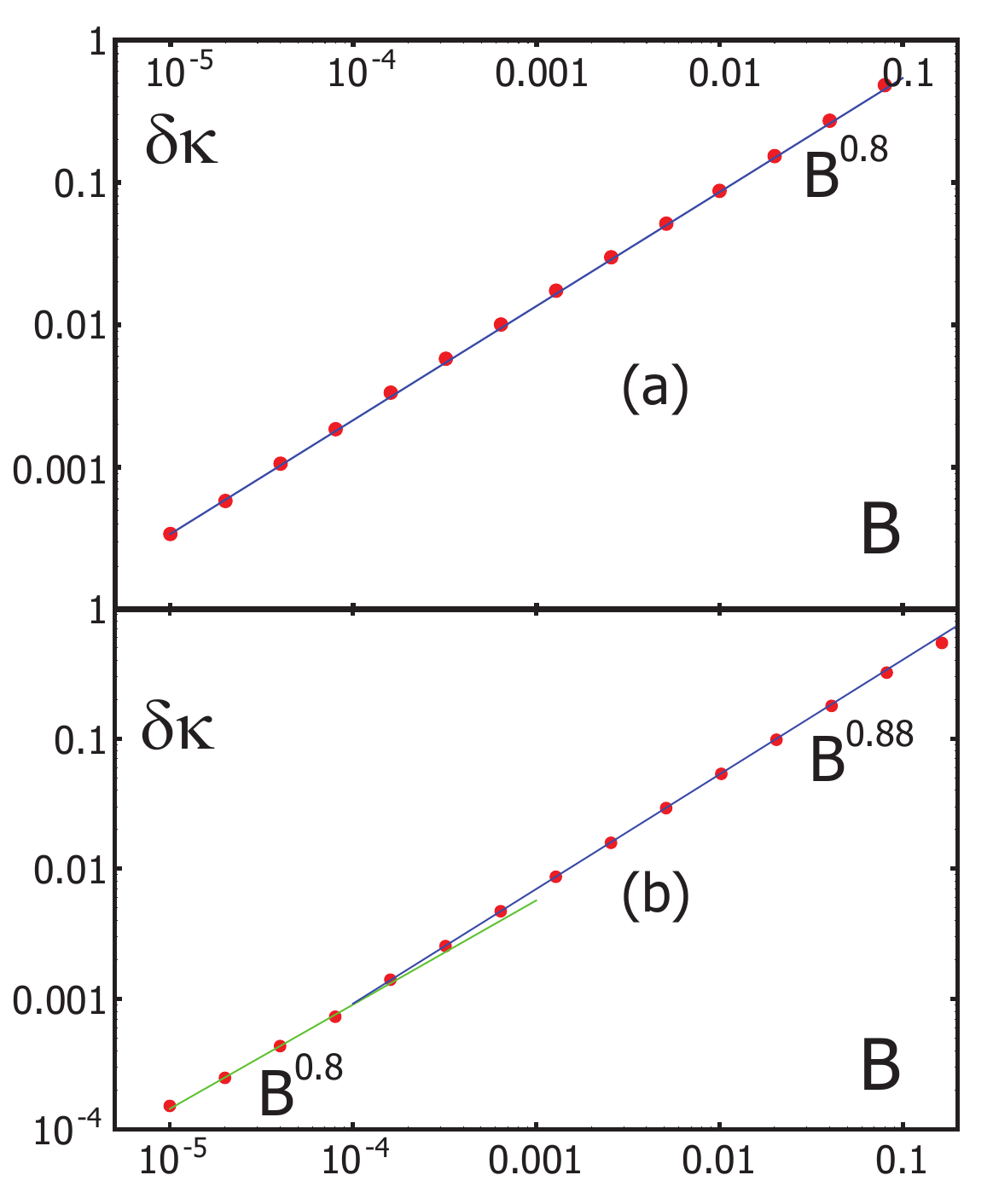}
\caption{Change in the inverse correlation length, $\delta\kappa=\delta(1/\xi)$
induced by magnetic field. The upper panels correspond to the gapped
density states, the lower panels correspond to linear density states.
The left panels correspond to random scattering amplitude signs, the
right one+ to positive scattering. The absolute values of inverse
correlation length at $B=0$ in these cases are $\xi_{0}^{-1}=1.85,\,1.42$
for gapped and linear densities of states, $X=0.5$ and $\xi_{0}^{-1}=1.31,\,0.95$
for uniform and linear density of states for positive scattering ($X=0)$.
The results for the constant density of states (not shown) are very
similar to the ones for the linear density of states shown in lower
panels: they display large intermediate regime of power law behavior
with exponents $0.6$ and $0.9$ for $X=0.5$ and $X=0.0$ respectively. }

\label{fig:dKappa(B)} 
\end{figure}

At intermediate fields one often observes a slightly different power
law 
\begin{equation}
g(B)=\frac{\Delta\xi(B)}{\xi(0)}=\pm D_{\pm}\left(\frac{B\xi^{2}}{\Phi_{0}}\right)^{\alpha'}\label{eq:NonUniversalCor}
\end{equation}
with a different exponent and pref actors, $\alpha'\approx0.5,\,0.6,\,0.64$
for the scattering of random signs with gaped, linear and uniform
densities of states respectively. For these densities of states the
pref actors are $D_{+}\approx0.11,\,0.22,\,0.30$. The value of $D_{+}$
for the gaped density of states is in agreement with the numerical
simulations of the previous workers \cite{Medina1990,Zhao1991}. Note
that the value of $D_{+}$ for the uniform density of states is roughly
three times larger that for the gapped one. This makes it possible
to observe large negative magnetoresistance experimentally as we discuss
in section \ref{sec:Review-of-the}. These statements are illustrated
in Figs.~\ref{fig:dKappa(B)}. The scaling dependence with the exponent
$\alpha'\approx0.6$ was observed previously in a number of works\cite{Medina1996,Zhao1991}
in which insufficient system sizes prevented the observation of the
asymptotic behavior.

We now give qualitative arguments that reproduce the observed scaling
behavior of the change in the localization length explained above. 

\begin{figure}[ptb]
\includegraphics[width=0.5\textwidth]{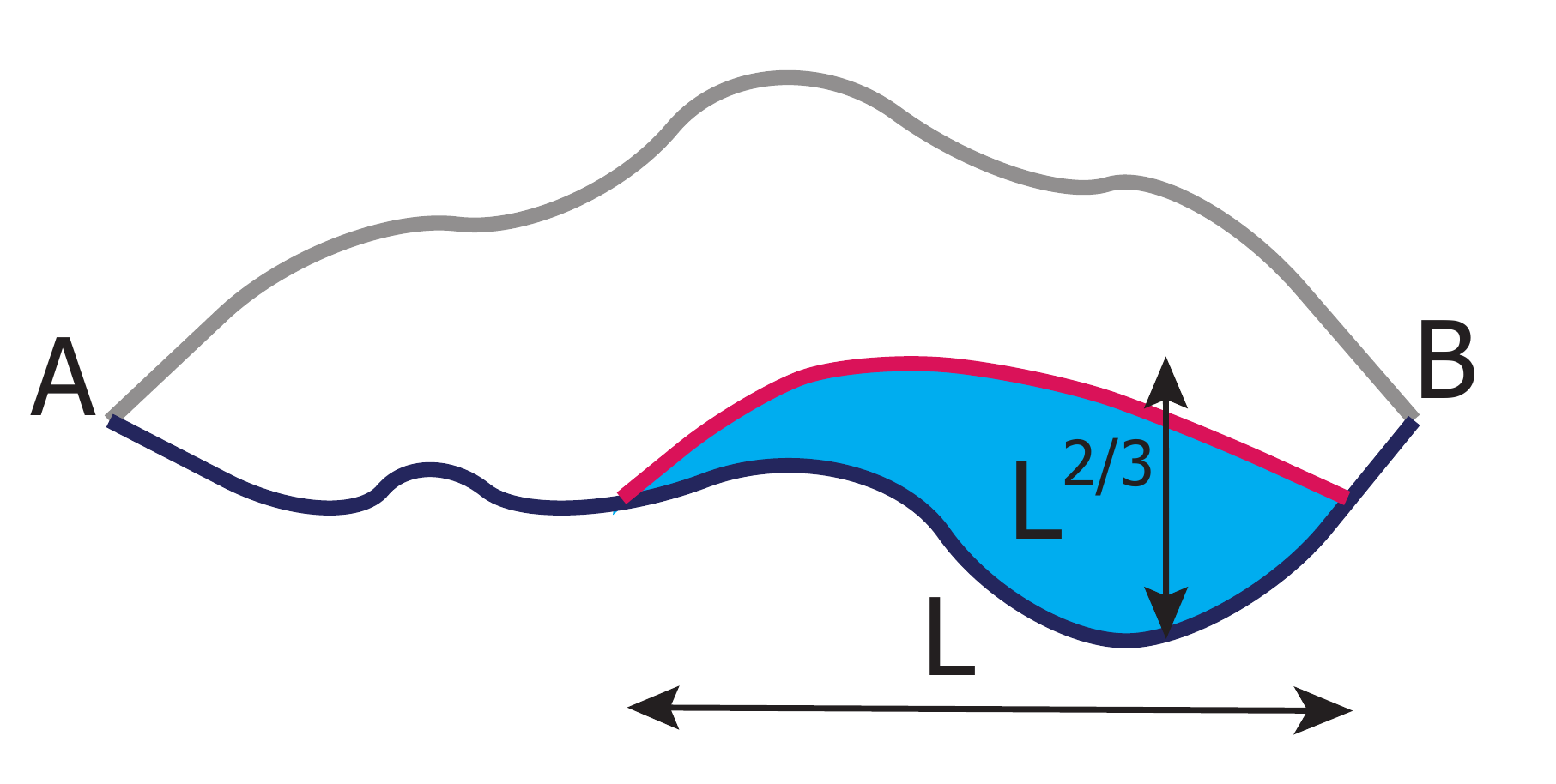}
\caption{Directed polymer picture}

\label{fig:EffectOfMagneticField} 
\end{figure}

As we have shown in section \ref{sub:Statistics-of-the}, the problem
of electron tunneling belongs to the same universality class as the
problem of directed polymers. In particular, the typical tunneling
action varies from one path to another by the amount that scales as
$\Delta F\propto L^{1/3}.$ This means that the tunneling from point
$i$ to $f$ is dominated by a narrow bundle of paths as shown in
Fig. \ref{fig:EffectOfMagneticField}. The width of this bundle does
not increase with the length of the path, so the magnetic field has
very little effect on the tunneling in this approximation. Another
bundle of paths that differs from the dominant one at scale $L$ has
action that is typically larger than that of the dominant path by
$\Delta F\propto L^{1/3}$, so its amplitude is exponentially suppressed
by $\exp\left[-c(L/a)^{1/3}\right]$. Here $a$ is the mean free path
of the electron (lattice spacing in the case of numerical simulations).
This leads to an exponentially small effect of magnetic field. However,
because the difference of the actions between two paths is a random
variable itself, with probability $p\propto L^{-1/3}$ two actions
differ only by the amount of the order of unity. If all scattering
amplitudes are positive, the change in the interference caused by
magnetic field decreases the total amplitude by the factor of the
order of unity, provided that the flux through the loop formed by
these two paths is of the order of the flux quantum. Because the transverse
direction scales as $Y\sim a(L/a)^{2/3}$ the interference becomes
relevant at scales 
\begin{equation}
BL^{5/3}a^{1/3}\sim\Phi_{0}\label{eq:BLa}
\end{equation}
with probability $p\sim(L/a)^{-1/3}.$ The resulting decrease of the
wave function implies that the typical inverse correlation length
increases by 
\[
\delta\xi^{-1}\sim a^{1/3}/L^{4/3}\sim(B/\Phi_{0})^{4/5}a^{3/5}
\]

Repeating the same arguments for the case of the amplitudes of the
random signs and using the fact that the signs of two paths that contribute
to the interference are random (cf. discussion after equation (\ref{eq:A_Gamma(B)})),
we get the same dependence on magnetic field but with the opposite
sign: the inverse correlation length is decreased by magnetic field. 

All these conclusions are valid in the limit of long scales where
$\Delta F\gg1$. In the intermediate regime, in which $\Delta F\lesssim1$
the probability that two paths interfere is of the order of unity
resulting in the scaling dependence of $\delta\xi^{-1}$on the field
with the exponent $\alpha'=3/5$. Looking at the numerical results
for the scaling dependence of $\Delta F$ shown in Fig. \ref{fig:Delta kappa distribution for linear DOS positive signs}
we see that it remains of the order of unity for $L\lesssim10^{2}$
which translates into the field $B\gtrsim10^{-3}$ in rough agreement
with the numerical results shown in Fig. \ref{fig:dKappa(B)}. 

\begin{figure}[ptb]
\includegraphics[width=0.5\textwidth]{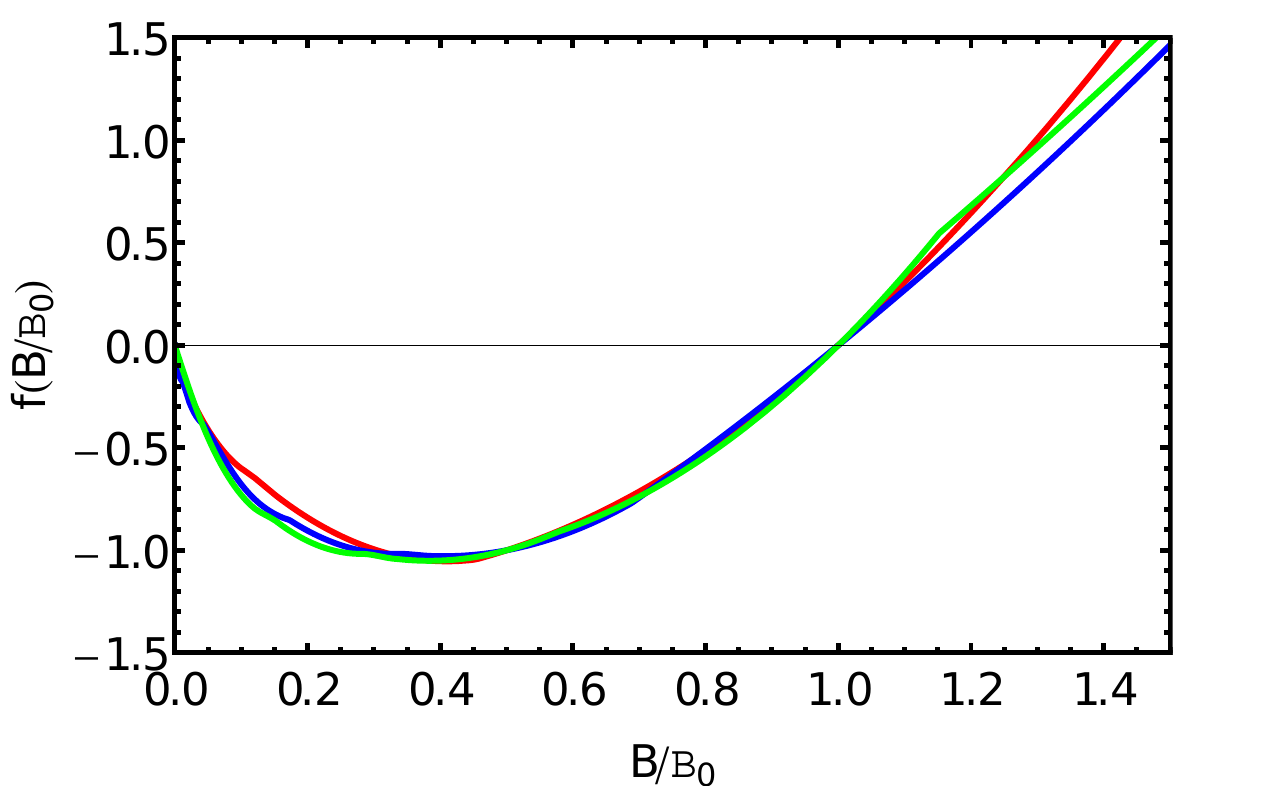}
\caption{Universal behavior of the increment of the inverse localization length
as a function of the field for small concentration of negative $\epsilon_{ij}$
and linear density of states. Different curves show $\delta\xi^{-1}$for
different concentrations $X=0.02,\,0.08$ and $0.16$ rescaled in
both vertical and horizontal directions: $\delta\xi^{-1}=\delta\xi_{0}^{-1}\chi(B/B_{0})$.
The characteristic value of the field scales with $X$: $B_{0}\propto X^{\beta}$
with $\beta\approx2.8$ (insert). Very small values of the field imply
that negative correction of $\delta\xi^{-1}$ wins over positive only
when signs are completely randomized, even a small correlation between
the signs of the amplitude is sufficient to result in the positive
correction. }

\label{fig:Universal kappa at small fields} 
\end{figure}

The behavior of the correlation length is given by the simple scaling
equations (\ref{eq:UniversalCor},\ref{eq:NonUniversalCor}) only
in the limit of completely random and positive amplitude signs. In
the case of a small concentration of negative scatterings one expects
a more complicated behavior. Large fields affect amplitude at short
scales. At these scales the rare negative scatterings have small effect
on the amplitude sign, so at large fields the inverse localization
length is increased by the field, similarly to the case of positive
scattering amplitudes. In contrast, at large scales relevant for small
fields the amplitude sign becomes completely random, so at small fields
one expects a negative correction to $\delta\xi^{-1},$ similar to
a fully random sign case. As the field is increased, the sign of the
correction should change. Exactly this qualitative behavior is shown
by numerical simulations of the model (\ref{eq:A_ij(B)}) with a small
concentration of scatterers with negative amplitudes. Our results
shown in Fig. \ref{fig:Universal kappa at small fields} display universal
behavior of $\delta\xi^{-1}(B/B_{0})$. The characteristic field $B_{0}$
scales, as expected, with concentration $X$: $B_{0}\propto X^{\beta}$,
however, the value of exponent $\beta\approx2.8$ is sufficiently
larger than one would expect from the scaling behavior of $L(x)\propto1/x$
obtained in section \ref{sub:The-sign-transition.}: $\beta_{expected}\approx1.6$.
We do not have a satisfactory explanation of this discrepancy. We
only note that very small values of $B_{0}$ found numerically imply
that even a small amount of sign correlations is sufficient to result
in the positive $\delta\xi^{-1}$. This is not so surprising because
positive increment of $\delta\xi^{-1}$, although given by the same
scaling dependence, is order of magnitude larger than the negative
one (cf. right and left panels of Fig. \ref{fig:dKappa(B)}). 

The scaling dependence (\ref{eq:UniversalCor}) is non-analytic in
$B$, so it should dominate over other sources of corrections to the
localization radius at $B\rightarrow0$. In the electron hopping problem
the largest scale, $r_{hop}$, for the coherent electron tunneling
is set by temperature (\ref{eq:r_VRH}). The non-analytic behavior
predicted by (\ref{eq:UniversalCor}) takes place provided that the
scale $L$ given by (\ref{eq:BLa}) is less than $r_{hop}(T)$: 
\[
(\Phi_{0}/B)^{3/5}a^{1/5}<\xi\left(\frac{T_{0}}{T}\right)^{\zeta}.
\]

\begin{figure}[ptb]
\includegraphics[width=0.5\textwidth]{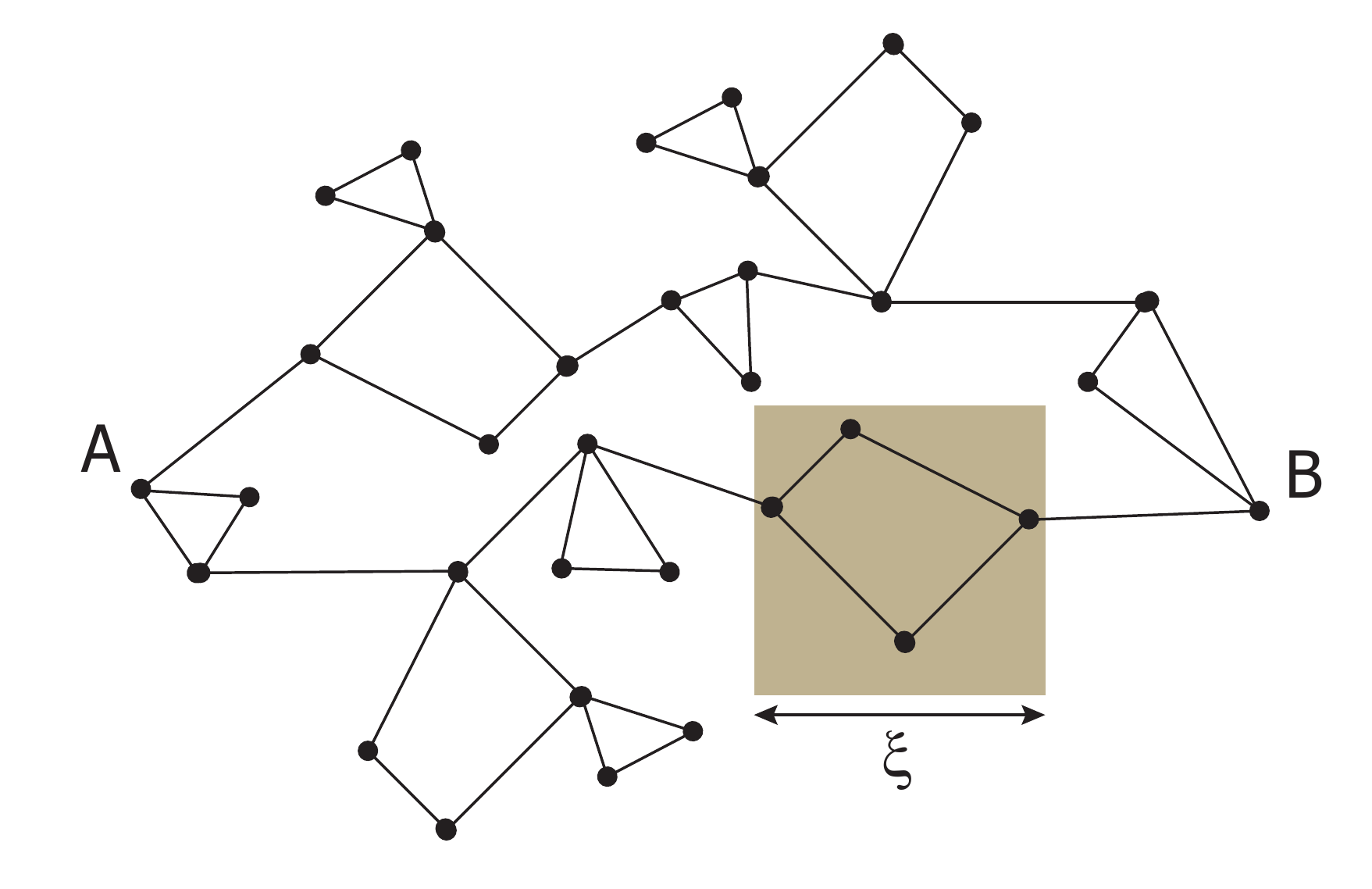} \caption{Quantitative picture of the tunneling paths in the vicinity of metal
insulator transition when $\xi>a_{B}$. The path may contain return
loops at short scales (of the order of $\xi)$ but at longer scales
the electron moves only in one directions. We expect that the problem
is mapped onto directed polymers at scales larger than $\xi$, so
that small magnetic fields $B\xi^{2}\lesssim\Phi_{0}$ are expected
to have the same effect on the resistivity as the in the strongly
localized regime. }

\label{fig:NearMItr} 
\end{figure}

In the discussion of the hopping transport we have assumed the strongly
localized regime in which the electron wave function is localized
at the scales of the order of the Bohr radius, $a_{B}$ of a single
impurity. However, all our qualitative conclusions should also hold
when localization length is larger, $\xi>a_{B}.$ In this case the
electrons tunnel from one area to another as shown in Fig.~\ref{fig:NearMItr}.
The loops of the tunneling paths are allowed inside individual areas,
but not between them. In this regime one expects to observe large
non-analytic dependence of the localization length on magnetic field
given by (\ref{eq:UniversalCor},\ref{eq:NonUniversalCor}) at low
fields $B\xi^{2}\lesssim\Phi_{0}$. These universal corrections adds
to the effect of magnetic field coming from the scales shorter than
$\xi$ that may be found from the renormalization group approach.
These corrections are of the order of $\delta\xi/\xi\sim(B\xi^{2}/\Phi_{0})^{2}$
and thus are negligible compared to the effects (\ref{eq:UniversalCor},\ref{eq:NonUniversalCor})
coming from the longer scales at low fields. They can, however, contribute
significantly to the total variation of the magnetoresistance at large
fields. 

\subsection{Magnetoresistance in variable range hopping regime. \label{sub:Prediction-for-magnetoresistance}}

The results (\ref{eq:UniversalCor},\ref{eq:NonUniversalCor}) for
the $\xi(B)$ dependence can be converted into magnetoresistance provided
that the induced change of the localization length is small $\delta\xi\ll\xi,$
but the resulting change in the hopping amplitude is exponentially
large leading to resistance variations $\ln\left(\rho\left(0\right)/\rho\left(B\right)\right)\gg1$.
In this case one can neglect other contributions to the variation
of the hopping probability (that we discuss below) so that magnetoresistance
is given by

\begin{equation}
\ln\left[\frac{\rho\left(B\right)}{\rho\left(0\right)}\right]\approx\left[2\zeta\left(\frac{T_{0}}{T}\right)^{\zeta}\right]\frac{\delta\xi}{\xi}\label{eq:ln_rho(B)_approx}
\end{equation}
Combined with the $\xi(B)$ dependence discussed in section \ref{sub:Predictions-of-single}
this equation gives the magnetoresistance at moderate fields, so that
$B\xi^{2}\lesssim\Phi_{0}$ but $\ln\left(\rho\left(0\right)/\rho\left(B\right)\right)\gg1$. 

At large magnetic fields $B\xi^{2}\gtrsim\Phi_{0}$ the equation (\ref{eq:ln_rho(B)_approx})
remains valid but the localization length dependence on magnetic field
is due to short scales and is non-universal. For a granular metal
the localization length is roughly equal to the grains size $r_{0}$;
because the magnetic field has no effect at scales shorter than $r_{0}$,
$\delta\xi(B)$ dependence saturates at $B\xi^{2}\lesssim\Phi_{0}$.
In contrast, in the case of a weakly disordered non-interacting 2D
metal with $kl_{tr}$>1 one expects\textbf{\cite{Lee1985}} strong
dependence on magnetic field. Indeed, in this case the localization
length is exponentially large $\xi\left(0\right)\sim l_{tr}\exp\left(k_{F}l_{tr}\right)$
in the absence of magnetic field, $l_{tr}$ being electron mean free
path. The conventional renormalization group analysis\textbf{\cite{Lee1985}}
gives $\delta\xi(B)/\xi(0)\sim(B\xi^{2}/\Phi_{0})^{2}$ at $B\xi^{2}<\Phi_{0}$,
so one expects corrections of the order of unity at $B\xi^{2}\approx\Phi_{0}$.
At larger fields ($Bl_{tr}^{2}\sim\Phi_{0}$) the localization length
increases exponentially to $\xi(B)\sim l_{tr}\exp\left(k_{F}l_{tr}\right)^{2}$.
At even larger fields one expects the appearance of the quantum Hall
regime and a pseudometallic behavior.\cite{Spivak2010} The presence
of electron-electron interaction can lead to even larger variety in
the localization length dependence on magnetic field at high fields.

The computation of $\xi(B)$ dependence in section \ref{sub:Predictions-of-single}
translates into the predictions for magnetoresistance (\ref{eq:ln_rho(B)_approx})
only in the asymptotic regime of large magnetic field at which $\ln\left(\rho\left(0\right)/\rho\left(B\right)\right)\gg1$.
There are at least two reasons why it is important to study the magnetoresistance
in the opposite limit of low magnetic field. 

First, because it is difficult to measure large resistances, the parameter
$r/\xi\lesssim15$ cannot be very large, so the condition is $\ln\left(\rho\left(0\right)/\rho\left(B\right)\right)\gg1$
is satisfied only in a limited range of fields. As we show below,
the power law dependence of $\ln\left(\rho\left(0\right)/\rho\left(B\right)\right)$
extends somewhat in the regime if $\ln\left(\rho\left(0\right)/\rho\left(B\right))\right)\lesssim1$
which makes the observation of this dependence more realistic.

Second, many data show that the magnetoresistance often changes sign
in small fields. As we discuss in more detail below, this sign change
agrees with the theoretical expectations. For instance, if the scattering
amplitudes are mostly positive ($P_{-}\ll1$) the localization length
at large fields becomes shorter (see section \ref{sub:The-sign-transition.})
and magnetoresistance is positive. However, at small fields it may
change its sign and become negative. This change in the sign of the
magnetoresistance can be due to the change in the sign of the correction
to the localization length discussed in section \ref{sub:The-sign-transition.}or
to another effect at short scales that we discuss below. Generally,
the theoretical predictions in this regime are less universal.

At small magnetic field the accuracy of the approximation $M_{ij}\sim A_{ij}$
becomes insufficient because it overestimates contributions to the
hopping rate (\ref{eq:W_ij}) from the impurity configurations in
which the partial amplitudes $A_{\Gamma}\left(0\right)$cancel each
other in the absence of magnetic field, so that the value of $A_{if}(0)\approx0$.
For these configurations a small magnetic field changes $\ln A_{if}$
dramatically. For a finite probability density of $A_{ij}(0)=0$ the
magnetic field dependence of $\overline{\ln A(B)}$ becomes a non-analytic
function of $B$: $\overline{\ln\left[A(B)/A(0)\right]}\propto|B|$.
\cite{Nguen1985,Shklovskii1990a} Similarly to the qualitative discussion
of $\xi(B)$ dependence in section \ref{sub:Predictions-of-single}
this non-analyticity can be demonstrated in the case when propagation
amplitude is due to the interference between just two paths: $A_{if}=A_{1}+A_{2}\approx0$
with random $A_{1}$and $A_{2}$. In this model case the typical amplitude
in magnetic field becomes 
\begin{equation}
\overline{\ln\left|\frac{A(B)}{A(0)}\right|}=\int dA_{1}dA_{2}\ln\left|A_{1}-A_{2}e^{i\phi}\right|\sim\left|\phi\right|\label{eq:ln(A(B))}
\end{equation}
where $\phi\propto B$ is the phase difference induced by the magnetic
field. Here and below we denote by bar the averaging over the impurity
configurations. Because the probability density of $A_{ij}(0)=0$
is finite at any concentration of scatterers with $\mu_{i}<0$, the
typical amplitude always grows at small fields. This however does
not \emph{always} translate into negative magnetoresistance. 

The crucial difference between the amplitude $A_{ij}$ and the hopping
rate (\ref{eq:W_ij}) is that the latter is the sum of the positive
rates due to phonons with different $q$ directions. As a result,
the probability density to find $W_{if}=0$ is zero, and at small
$B$ the magnetorsistance is proportional to $B^{2}$.

In order to find the values of the crossover fields we note that in
the limit of low temperatures at which $qr_{ij}\ll1$ the exponential
in (\ref{eq:M_ij}) can be approximated by the first non-zero term:
\begin{equation}
M_{ij}(\overrightarrow{q})\sim\int d\vec{r}\psi_{i}^{\dagger}\left(\vec{r}\right)\psi_{j}\left(\vec{r}\right)\overrightarrow{q}\vec{r}.\label{eq:M_ij(q)}
\end{equation}

The main contribution to the matrix element $M_{ij}$ comes from the
components of the phonon wave vector $\overrightarrow{q}$ which is
parallel to $\vec{r}_{ij}$. In the leading approximation we can neglect
the contributions from the phonons with momenta in other directions.
In this approximation the hopping probability (\ref{eq:W_ij}) is
controlled by the matrix element $M_{ij}(q\hat{r}_{ij})$ $\hat{r}_{ij}=\overrightarrow{r_{ij}}/r_{ij}$.
This matrix element has the same statistical properties as the amplitude
$A_{if}$, so the reasoning resulting in (\ref{eq:ln(A(B))}) applies
and $\overline{\ln\left|M\left(B,\overrightarrow{q}\right)/M\left(0,\overrightarrow{q}\right)\right|}\sim\left|B\right|$.
The subleading processes in which the hopping (\ref{eq:W_ij}) is
due to phonons with momenta perpendicular to $\vec{r}_{ij}$ cut off
the non-analytic behavior of $\overline{\ln W(B)}$ at very small
fields. 

Combining this result with the effect of $\xi(B)$ dependence discussed
in section \ref{sub:Predictions-of-single} that takes place at large
scales at which the flux through the typical loop is larger than the
flux quantum, $Br^{5/3}\xi^{1/3}>\Phi_{0}$, we get three regimes
of the $\overline{\ln M(B)}$ dependence for $B\xi^{2}<\Phi_{0}$: 

\begin{equation}
\ln\frac{\rho(0)}{\rho(B)}=\overline{\ln\frac{W_{if}(B)}{W_{if}(0)}}\sim\left\{ \begin{array}{cc}
(B/B_{0})^{\alpha}\gtrsim1 & B>B_{0}\\
|B|/B_{0}\lesssim1 & B_{0}>B>B_{*}\\
B^{2}/(B_{*}B_{0})\ll1 & B<B_{*}
\end{array}\right.\label{eq:ln(W(B)/W(0))}
\end{equation}
where $B_{0}=\Phi_{0}/r^{5/3}\xi^{1/3}.$ As we saw in section \ref{sub:Predictions-of-single}
the transverse deviations of the typical path scale as $r_{\perp}\sim r^{2/3}\xi^{1/3}$.
This allows us to estimate the contribution to the average (\ref{eq:W_ij})
from phonons with $q\perp r$: $W_{\perp}\sim(\xi/r)^{2/3}W$. Repeating
the arguments that led to (\ref{eq:ln(A(B))}) we get
\begin{equation}
\overline{\ln\frac{W(B)}{W(0)}}=\int dW_{\parallel}\ln\left[W_{\parallel}+W_{typ}\phi^{2}+W_{\perp}\right]\label{eq:ln(W(B)/W(0))b}
\end{equation}
that results in the dependence (\ref{eq:ln(W(B)/W(0))b}) with $B_{*}=\Phi_{0}/r^{2}$. 

The qualitative estimates show that while the regime of non-analytic
dependence is relatively wide ($B_{0}<B<\Phi_{0}/r^{2})$ the regime
of the linear dependence is narrow. We note that the estimates of
$B_{*}$and $B_{0}$ neglect the numerical coefficients that might
be important.

The discussion above and the result (\ref{eq:ln(W(B)/W(0))}) assumed
that the system is deep in the sign disordered phase in which signs
of all amplitudes are completely random. If the scattering amplitudes
are mostly positive $P_{-}\ll1$ the signs of the amplitudes become
random only at large scales. It implies that the system may be in
the sign ordered phase at characteristic scales set by magnetic field.
In this case the magnetoresistance at largest fields is positive in
contrast to (\ref{eq:ln(W(B)/W(0))}), while at small $B$ it is quadratic
in $B,$ so it can be both positive and negative depending on the
value of $P_{-}\ll1$. 

\begin{figure}
\includegraphics[width=0.5\textwidth]{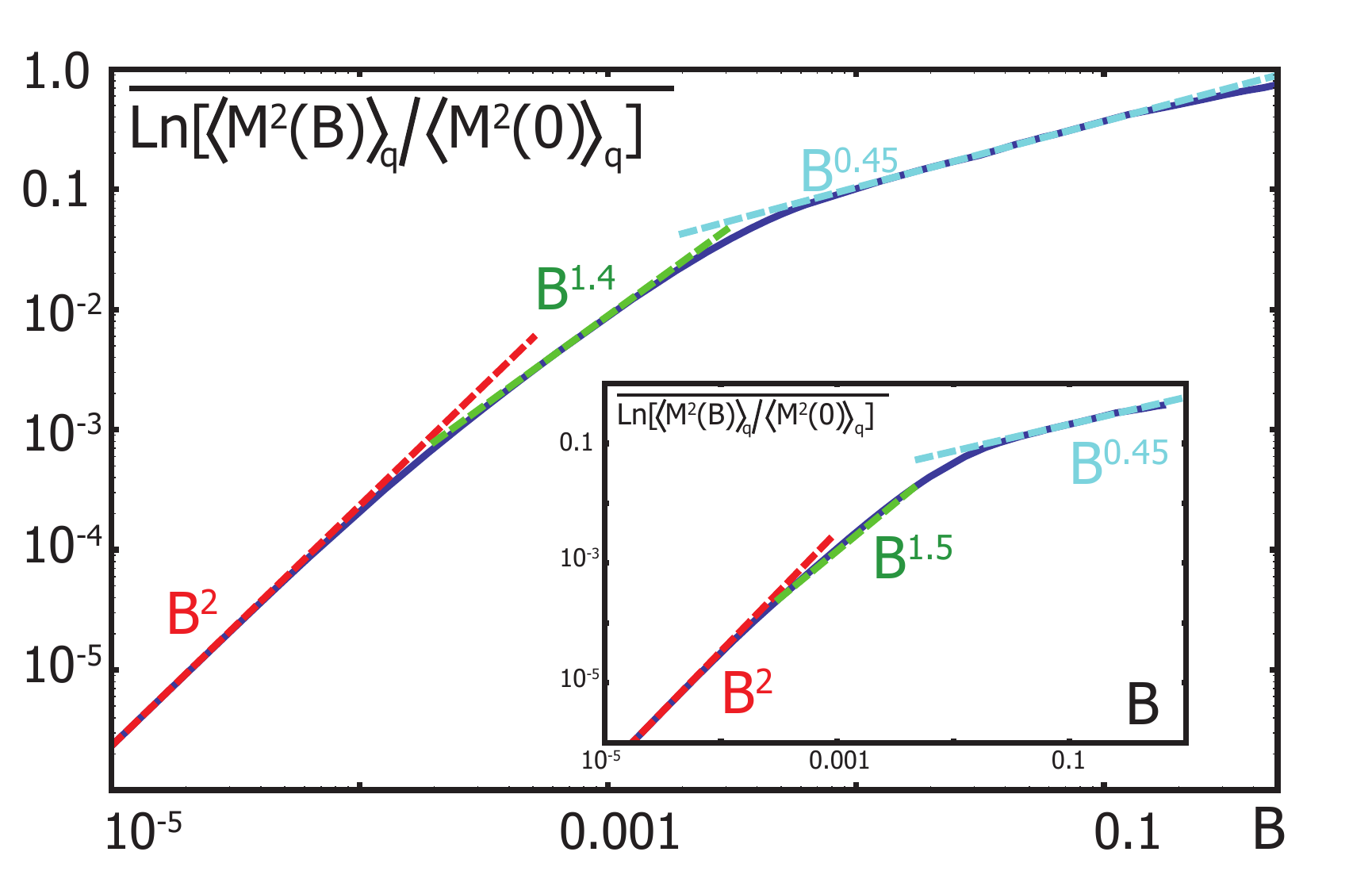}

\caption{Phonon matrix element as a function of magnetic field at long and
moderate scales. The main panel shows field dependence of the matrix
element for relatively long hops for which corresponding to $r/\xi\approx8.0$
One observes a very significant (two decades) regime of the pseudo-universal
scaling dependence associated with the localization length dependence
(\ref{eq:NonUniversalCor}). At shorter scales (corresponding to $r/\xi\approx6$
shown in the insert) the scaling regime shrinks. In both cases the
regime of analytical dependence ($B^{2})$ is limited to very small
fields. }
\label{fig:MatrixElement}
\end{figure}

\begin{figure}
\includegraphics[width=0.5\textwidth]{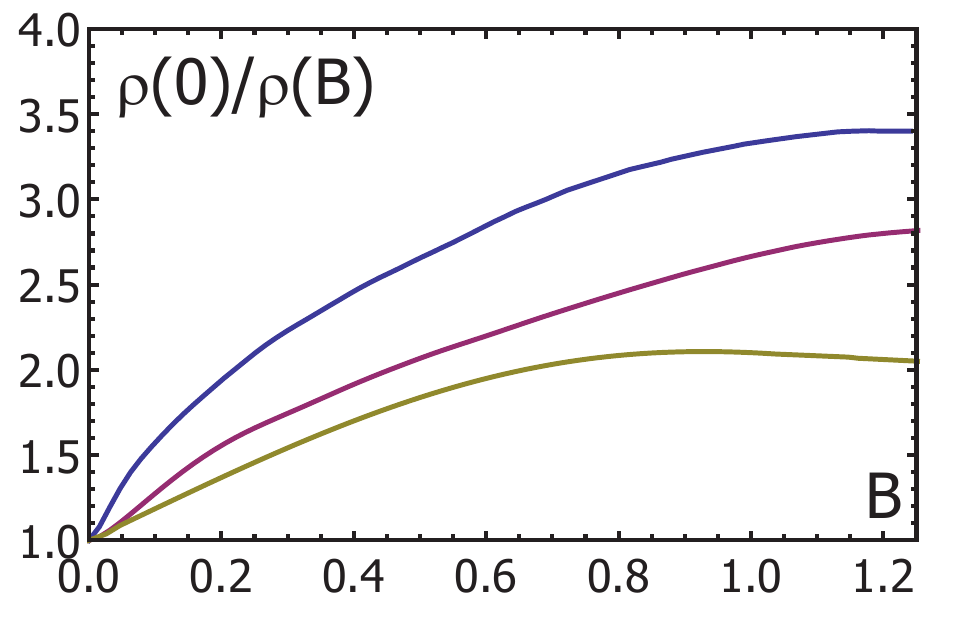}

\caption{Magnetoconductance as a function of magnetic field for different values
of the matrix element at zero field corresponding to $R_{\square}/R_{Q}\approx\;5\,10^{5},\;1.0\,10^{4},\;1.0\,10^{3}$.
For small matrix elements (large resistances) the behavior at low
fields can be approximated by a power law $\sigma(B)/\sigma(0)\approx B^{a}$
with $a\approx0.5-0.6$. The regime of very small magnetic fields
is hardly observable on the linear scale of the plot even for smallest
resistances.}
\label{fig:Magnetoconductance}
\end{figure}

To check the validity of (\ref{eq:ln(W(B)/W(0))b}) for realistic
parameters we have performed the numerical computation of the matrix
elements. We did not attempt a full computation of the matrix element
and its averaging over the distribution of $r_{ij}$ that characterize
the percolating cluster. Instead, we computed matrix element for the
characteristic $r_{ij}$ and averaged over different direction of
$q$. Because the results do not change qualitatively when $r$ is
increased by a factor of $2$, we believe that they reproduce faithfully
the dependence of the magnetoresistance:

\begin{equation}
\ln(\rho(0)/\rho(B))=\overline{\ln\left[\left\langle M^{2}(B)\right\rangle _{q}/\left\langle M^{2}(0)\right\rangle _{q}\right]}\label{eq:ln(rho(B)/rho(0))_approx}
\end{equation}
where angular brackets denote averaging of the directions of $\vec{q}$.
The result of our numerical simulations for the case of uniform density
of states $P(\epsilon)=(1/2)\theta(1-|\epsilon|)$ is shown in Fig.
\ref{fig:MatrixElement} for two typical distances: $r/\xi\approx8$
and $r/\xi\approx6$. In both cases one observes a large regime of
the pseudo-universal behavior $\ln(\rho(0)/\rho(B))\sim B^{\alpha}$
with $\alpha\approx0.5$ that is due to the non-universal corrections
to the localization length (\ref{eq:NonUniversalCor}). At larger
$r/\xi\gtrsim8$ one observes the gradual appearance of the transient
linear dependence in magnetic field in agreement with the expectations
(\ref{eq:ln(W(B)/W(0))b}). Figure \ref{fig:Magnetoconductance} shows
expected magnetoconductance at different typical values of $r/\xi$
converted into expected values of the resistances. 

\subsection{Beyond the single particle model. \label{sub:Beyond-single-particle}}

So far in our discussion we have ignored the many body effects due
to electron-electron interaction. Generally one expects that electron
correlations play much bigger role in the hopping regime than in the
metallic regime. In this subsection we briefly discuss their role,
and the conditions under which the single particle results obtained
above are valid. 

At low temperatures the electron sites with $\epsilon_{\alpha}<0$,
(and $\mu_{\alpha}<0$), are occupied by electrons, while the sites
with $\epsilon_{\alpha}>0$ are empty. Tunneling between initial and
final states may be viewed as a virtual process in which the electron
hops through the intermediate localized states. Depending on the ratio
between the electron-electron interaction and the density of states
at the Fermi energy in the impurity band, these localized states can
be singly and doubly occupied. The spins in the singly occupied states
interact via the exchange interaction, $J$. Although the detailed
theory of the disordered electron systems does not exit, three obvious
limiting cases are clearly possible. In the first case, the interaction
between electrons is large, so the majority of sites are singly occupied,
and the resulting spin system might form a $S=1/2$ spin glass at
low temperatures and a paramagnet at high temperatures. The low temperature
spin glass state breaks the time reversal symmetry, it might be collinear
or isotropic depending on the anisotropy of the exchange couplings.
Although logically possible, neither collinear nor isotropic states
were observed experimentally, probably because quantum spin fluctuations
are too large for spin $1/2$. The alternative (second case) is that
each spin forms a singlet with another spin to which it is coupled
by the strongest interaction \cite{Bhatt1982}. This state does not
break the time reversal symmetry. Finally, in the limit of small interaction
the majority of states are doubly occupied (third case). Both the
second and third cases are characterized by zero average spin on each
site. 

In all cases the segments of the tunneling path where electrons travel
through occupied sites may be viewed as a tunneling of a hole moving
backwards through occupied states as it is schematically shown in
Fig.~\ref{fig:HoppingViaLocalizedStates}. In the interacting system
this process may lead to the creation of many body excitations in
the final state that destroy the coherence between hopping amplitudes
$A_{\Gamma}$ along different paths $\Gamma$. When it does not happen,
the tunneling may be described by the equation (\ref{eq:A_ij(0)})
with renormalized hopping amplitudes and energies $\epsilon_{a}.$ 

We now discuss the tunneling interference in different electron states
in more detail. We start with a state in which all sites are single
occupied. At high temperatures the resulting spins form a paramagnet,
so the final spin state formed after the charge transport along different
paths $\Gamma$ are generally different, and do not coincide with
the initial state. In this state the corresponding amplitudes $A_{\Gamma}$
do not interfere. In this situation one expects no orbital effects
of the magnetic field on the charge transport. Application of magnetic
field can polarize the spin system, restoring the path interference.
Thus, in this case one expects that the polarization of the spin system
by the in-plane field results in a state characterized by a large
negative magnetoresistance with respect to the field perpendicular
to the plane, while application of a small perpendicular field in
the absence of in-plane one gives small or no negative magnetoresistance.
Large out-of-plane field (in the absence of in-plane field) has two
effects: it might polarize the spin system and cause orbital effects.
Thus, one expects a complicated behavior as a function of the out-of-plane
field. 

At low temperatures the spins may freeze in a spin glass state or
form a spin liquid. If the spins freeze in the collinear spin glass
state, the final states corresponding to two paths mostly coincide
and the interference reappears. In this situation the electron hopping
amplitude can be described by essentially the same equation (\ref{eq:A_ij(0)}).
Thus, one expects the same orbital effect of the magnetic field, as
discussed in section \ref{sub:Predictions-of-single}. 

The electron hopping becomes very different in the non-collinear spin
glass because the electron amplitudes acquire a non-trivial phase
factors due to spin non-collinearity which can be described by complex
scattering amplitudes $\mu_{a}$. We expect that magnetic field does
not affect the interference in this case and does not lead to orbital
magnetoresistance. However, the isotropic spin glass state is rather
unlikely to be realized in physical two and even three dimensional
glasses.\cite{Fischer1993} 

In contrast to the spin glass states, the spin singlets formed in
the second and third cases do not break the time reversal symmetry.
Thus, the scattering amplitudes in these situations remain real as
in the single particle model. At low temperatures the final states
formed after charge motion should coincide, so the interference between
different paths remains the same as it was in the one particle model
of section \ref{sub:Predictions-of-single}. 

We do not discuss here the effect of magnetic field on the spin configuration
which also affects the transport of charges. This discussion is beyond
the scope of this paper devoted to the orbital effects. We, however,
mention briefly possible scenarios in section \ref{sec:Review-of-the}
where we discuss the experiment that indicates that these effects
are important.

\section{Application to other physical systems.\label{sec:Application-to-other} }

The sign phase transition that appears for binary distribution of
scattering amplitudes discussed in section \ref{sub:The-sign-transition.}
can be observed in very different physical systems. Here we show that
it affects the physics of random classical magnets at high temperatures.
The simplest example is given by the Ising model on a cubic lattice
\begin{equation}
H=\sum_{ij}J_{ij}s_{i}s_{j}
\end{equation}
where $s_{i}=\pm1$ and the exchange interactions takes two values:
$J_{ij}=J_{0}>0$ with probability $(1-X)$ and $J_{ij}=J_{0}$ with
probability $X$ respectively. 

At high temperatures the susceptibility in this model 
\begin{equation}
\chi(\textbf{r}_{i},\textbf{r}_{f})=\langle s(\textbf{r}_{i})s(\textbf{r}_{f})\rangle=\sum_{\{s_{0}\}}s(\textbf{r}_{i})s(\textbf{r}_{f})\exp(-\frac{H}{T})\label{eq:IsHez}
\end{equation}
which is random quantity at large $r_{if}\gg1$. To show the existence
of sign phase transition in this quantity we notice that at $T\gg|J_{ij}|$
one can expand the exponent in (\ref{eq:IsHez}) and take into account
only directed paths between sites $i$ and $f$. The sum over directed
path is equivalent to solution of the recursion equation 
\begin{equation}
\chi_{km}=\chi_{k-1m}J_{k-1m}^{km}+\chi_{,m-1}J_{km-1n}^{km}\label{recurcionSpins}
\end{equation}
Here indices $(km)$ denote the site with coordinates $k,m$ on the
square lattice and $J_{k-1m}^{km}$ denote the bond connecting two
such sites. The recursion (\ref{recurcionSpins}) is very similar
to (\ref{eq:A_ij}) with binary distribution of $\epsilon_{ij}$,
so one expects that it shows the same sign transition as function
of concentration, $X$, of negative bonds. The only difference between
(\ref{recurcionSpins}) and (\ref{eq:A_ij}) is that in the former
the negative signs are associated with bonds and in the latter with
sites. This is similar to the difference between site and bond disorder
in percolation problem which is known to have very little effect.
Thus, we expect that at $r\rightarrow\infty$ the distribution function
of $\chi(r)$ exhibits the sign phase transition as a function of
$X$. At high temperatures the critical value $X_{c}$ is $T-$independent.
As the temperature is decreased, the sign correlations increase which
can lead to the formation of the sign ordered phase. This means that
the transition from spin disordered to spin ordered phase shifts to
larger $X$ at lower temperatures. Finally, at sufficiently low temperatures
the system might become a ferromagnet. At the transition point the
susceptibility (\ref{eq:IsHez}) decreases as a power law of $|r_{i}-r_{f}$|
and the sign correlations are long ranged whereas spin correlator
decreases exponentially. Thus, the transition to the sign ordered
state happens above the transition to a ferromagnet.

The staggered susceptibility is defined by $\tilde{\chi}(r)=(-1)^{n}\chi(r)$,
where $n$ is the number of steps in a direct path on square lattice
between the sites $0$ and $r$. Obviously it also exhibits a sign
phase transition. Thus, at high temperatures the sign disordered phase
is separated from the phases in which the sign of the susceptibility
is positive or alternating. At sufficiently low temperatures the system
freezes into a magnetically ordered or a the spin glass phase. The
spin glass phase may be sign ordered or disordered, the former corresponds
to the coexistence of ferromagnetic (or antiferromagnetic) and spin
glass order parameters. These conclusions are summarized by the phase
diagram shown in Fig.~\ref{fig:HoppingViaLocalizedStates}.

\begin{figure}[ptb]
\includegraphics[width=0.3\textwidth]{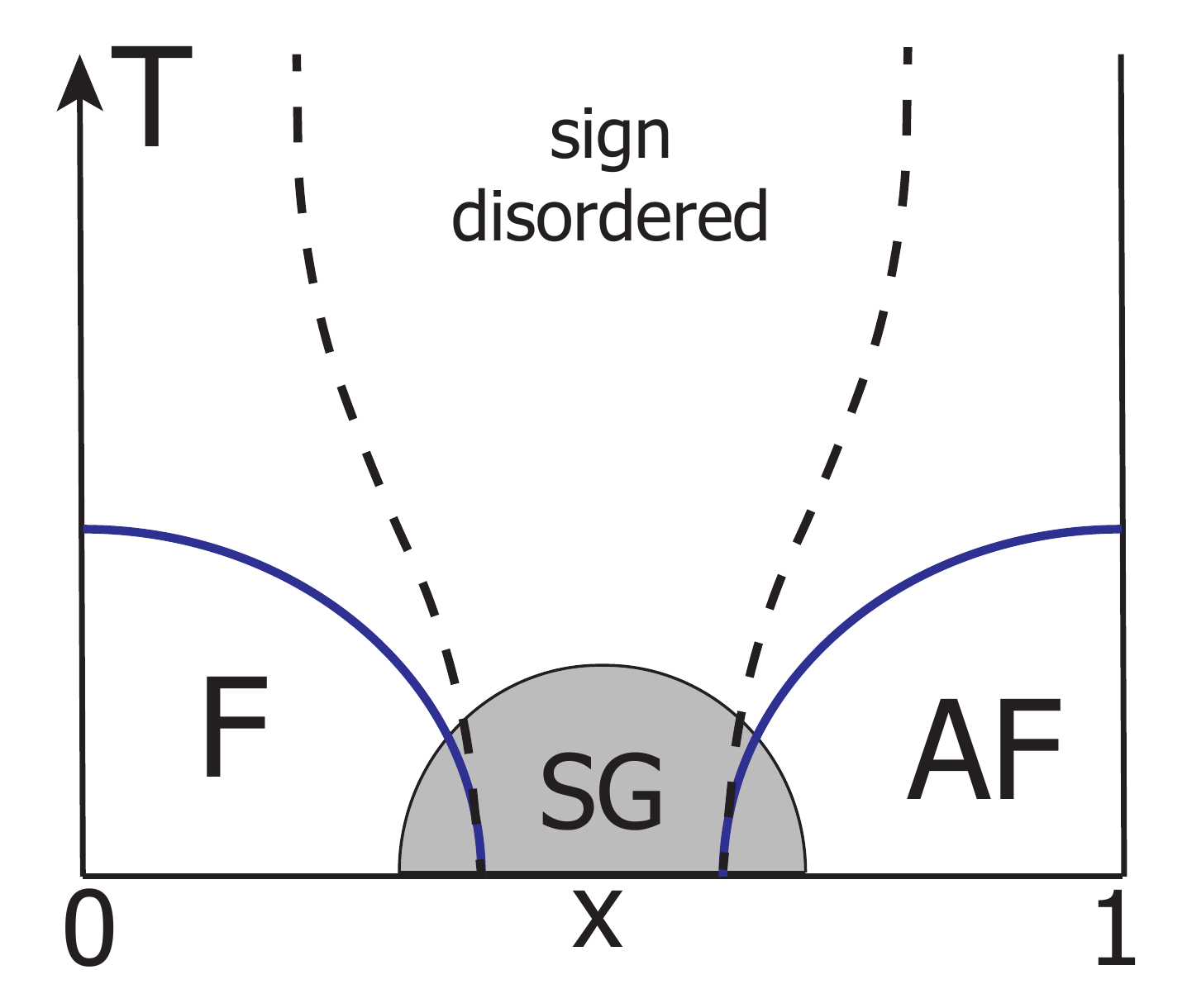}
\caption{Qualitative picture of a phase diagram of Ising spin glass. Dashed
lines separate sign ordered and sign disordered phases at high temperatures.
The spin glass phase (shown in gray) appears in dimension three and
higher. In two dimensions the spin system remains paramagnetic down
to lowest temperatures in the absence of the ferro- (or antiferro-)
magnetic long range order. }

\label{fig:Spinglass} 
\end{figure}

\section{Review of the experimental results and conclusions.\label{sec:Review-of-the} }

Theoretical expectations described in the previous sections can be
separated into the qualitative and quantitative predictions. Verification
of the qualitative prediction of the orbital mechanism of a large
negative magnetoresistance in the variable range hopping regime is
relatively simple: requires only measurements of the anisotropy with
respect to the parallel and perpendicular magnetic field. In contrast,
to verify quantitative predictions represented by (\ref{eq:UniversalCor},\ref{eq:NonUniversalCor})
would require stronger conditions $\ln\left[\varrho\left(0\right)/\rho\left(B\right)\right]>1$,
and $B\xi^{2}>\Phi_{0}$ . We are not aware of experiments on the
negative magnetoresistance where all these requirements were satisfied.
Below we discuss currently available data on large negative magnetoresistance
in the variable range hopping. 

We begin with the maximal value of the magnetoresistance observed
experimentally and expected theoretically. In our numerical simulations
we got the maximal value of $\delta\xi/\xi=0.2$ for the uniform (Mott
regime) and $\delta\xi/\xi=0.05$ for the linear in $\varepsilon$
(Efros-Shklovskii regime) density of states. The measurable values
of the resistance ($R\lesssim10^{11}\,\Omega$ ) correspond to $\left(T_{0}/T\right)^{\zeta}\lesssim15$.
Thus (\ref{eq:UniversalCor},\ref{eq:NonUniversalCor}) describe the
negative magnetoresistance whose value does not exceed $\varrho\left(0\right)/\rho\left(B\right)<30$
in the Mott regime, and is expected to be more moderate, $\ln[\rho\left(0\right)/\rho\left(B\right)]<1$,
in Efros-Shklovskii regime. This is in agreement with the fact that
in all works \cite{Savchenko1987,Jiang1992,Ovadyahu1990,Ovadyahu1995,Kravchenko1998,Spivak2010,Wang2006,Hong2011,Friedman1996,Mitin2007,Dynes1996}
where both the large negative magnetoresistance has been observed
and the temperature dependence of the resistance has been measured,
it followed Mott's law.

\begin{figure}
\includegraphics[width=3in]{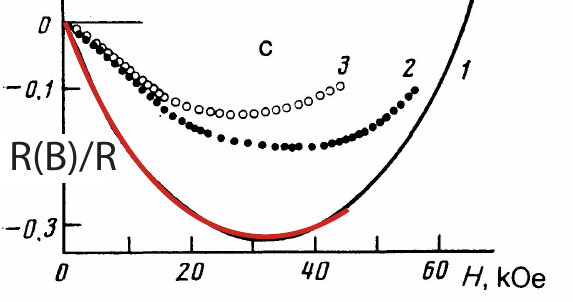}

\caption{Data from \textbf{\cite{Savchenko1987} }and their fit to the behavior
of (\ref{eq:UniversalCor}) expected for relatively small resistances
$R_{\square}/R_{Q}\approx10^{3}-10^{4}$ that used the matrix elements
computed in section \ref{sub:Prediction-for-magnetoresistance}. The
lower data curve correspond to the field perpendicular to the plane
of the sample. The middle data points show the effect of the field
in plane of the sample perpendicular to the direction of the current,
the highest data set to the field in the direction of the current.
The thick red line shows the theoretical expectations. The upturn
at the large fields is due to the effect of the field at small scales
where it modifies the hopping amplitude between the sites which was
not taken into account properly in the model. }
\label{fig:SavchenkoDataFit}
\end{figure}

Surprisingly, one of the most comprehensive studies of the negative
magnetoresistance in the variable range hopping regime in a two dimensional
material was done in the early work\cite{Savchenko1987} that studied
$Ge$-sopped $GaAs$ films. It observed a strongly anisotropic negative
magneto resistance, the largest one corresponding to the out-of-plane
field. The effect of the in-plane field can be accounted for by a
significant thickness of the film ($d_{eff}\approx30nm$). Moreover,
the in-plane negative magneto resistance was also anisotropic with
respect to the angle between the magnetic field and the current. Finally,
microscopic fluctuations of the resistance as a function of the magnetic
field in small samples was observed. These observations prove the
orbital nature of the effect. In this experiment the resistance of
the sample was $R_{\square}\lesssim30M\Omega$ at lowest temperatures
indicating that $r/\xi\lesssim5$. Accordingly the magnitude of the
negative magneto resistance remained moderate: $(\left(\rho\left(0\right)-\rho\left(B\right)\right)/\rho\left(0\right))_{max}\sim0.4$.
In Fig. \ref{fig:SavchenkoDataFit} we present results of our numerical
simulations of the equation (\ref{eq:ln(rho(B)/rho(0))_approx}) and
their comparison with the experimental data of \cite{Savchenko1987}.
The work\cite{Ovadyahu1990} observed negative magneto resistance
with similar amplitude and similar dependence on magnetic field in
thin films of polycrystalline $\mathrm{In_{2}O_{3-x}}$. 

A subsequent work \cite{Jiang1992}\textbf{ }on $\mathrm{GaAs/Al_{x}Ga_{1-x}As}$
disordered hetero junctions observed significantly larger negative
magneto resistance $\varrho(0)/\rho(B)\sim7$ . Strong anisotropy
of the negative magneto resistance has been observed indicating the
orbital nature of the effect. The magnetic field dependence of $\rho(B)$
in low fields $B\lesssim4\mathrm{\, T}$ where magneto resistance
is negative was roughly linear in coordinates $\ln\rho(B)$, $B^{1/2}$which
is in a good agreement with the dependence expected theoretically
(\ref{eq:ln(W(B)/W(0))b}) and shown in Fig. \ref{fig:Magnetoconductance}.
In these experiments the localization length varied between $\xi=25-100\,\mathrm{nm}$
for different gate voltages, so that $B\xi^{2}\sim\Phi_{0}$ occurs
at $B\sim4\mathrm{\, T}$. Generally one expects that the magneto
conductance should show a crossover to a different regime when $B\xi^{2}\sim\Phi_{0}$.
It is surprising that this crossover is not observed in the data.
On the other hand, this work and works discussed below give values
for the localization length $\xi$ extracted from the Mott law. This
procedure is prone to a number of uncertainties such as the value
of the density of states, the exact form of the temperature dependence,
etc, so the values of the localization length might be wrong by a
factor $2-5$ which would be sufficient to explain the absence of
the crossover in \cite{Jiang1992}. Similar large negative magneto
resistance ($\varrho(0)/\rho(B)\sim20$) of orbital nature has been
observed in polycrystalline $\mathrm{In_{2}O_{3-x}}$ films in\cite{Ovadyahu1995}.
The behavior of $\varrho(0)/\rho(B)$ in these experiments resembles
a small power of magnetic fields in a wide range of fields for all
fields, the quadratic behavior was observed only in very low fields
($B<0.2\,\mathrm{T}$), at which the relative change in the resistance
was very small $\delta R/R\ll1$ in agreement with the theoretical
expectations (cf. Fig. \ref{fig:MatrixElement} in which $B^{2}$
behavior appears at $\delta R/R\lesssim10^{-2}-10^{-1}$). 

The maximal value of magnetoresistance in works\textbf{ }\cite{Jiang1992,Ovadyahu1990,Ovadyahu1995}\textbf{
}is somewhat above the one expected theoretically for the films of
these resistances. For instance, the resistance of $\mathrm{GaAs/Al_{x}Ga_{1-x}As}$
films in work \cite{Jiang1992} implies that at the lowest temperature
the maximal value of $(T_{0}/T)^{1/3}\approx7$ for these films which
translates into the maximal expected value for $\rho\left(0\right)/\rho\left(B\right)\sim2-3$.
It is possible, however, that the largest fields studied in these
works correspond to the regime $B\xi^{2}\gtrsim\Phi_{0}$ in which
the magnetoresistance may continue to grow with $B$.

Huge effect of the transverse field on the conductivity ($\varrho\left(0\right)/\varrho\left(B\right)\gtrsim30$)
of high mobility silicon MOSFET was observed\cite{Kravchenko1998,Spivak2010}
at low carrier concentrations. Remarkably in these experiments the
large magnetoresistance in the transverse field appears only when
the spins are polarized by large in-plane field whilst low fields
result in an isotropic small and positive magnetoresistance. The latter
indicates its spin nature which is in agreement with the strong correlations
expected in this material. As discussed in section \ref{sub:Beyond-single-particle}
this implies the existence of the localized spins in the system that
suppresses the orbital effect of the magnetic field. Application of
a large in-plane field polarizes the spins making the path interference
possible, so that a transverse field added to the system leads to
a large negative magnetoresistance, as observed experimentally. Unfortunately,
the work\cite{Kravchenko1998}\textbf{ }did not study the temperature
dependence of the resistivity in these samples. It is likely that
the change of the sign of magnetoresistance observed in work\textbf{\cite{Wang2006}}
that studied the pregraphitic carbon nanofibers that obey Efros-Shklovskii
law is due to a similar mechanism. Unfortunately, this work did not
study the field anisotropy.

The paper\cite{Hong2011} reported a big negative magnetoresistance
($\varrho\left(0\right)/\varrho\left(B\right)\sim10$) of the H-doped
graphene, whilst in-plane field had practically no effect on the resistance.
The observed negative magnetoresistance can be interpreted as a large
change in the localization length: $\xi(B)/\xi(0)=4$ induced by $B=9\,\mathrm{T}$
field. These results cannot be compared directly with the universal
scaling dependence derived in this work because the large changes
in the localization length imply that $B\xi^{2}\sim1$ . One expects
that at lower temperatures the samples studied in this work should
exhibit large magnetoresistance at low fields associated with small
$\delta\xi/\xi$ but these data are not available. 

Finally, it is possible that negative magnetoresistance due to orbital
effect was observed in other materials as well but not studied in
any detail. For instance, the work\cite{Friedman1996} observed a
sharp (factor of $2$) drop of resistance in fields $B=1\,\mathrm{T}$
at $T=100\,\mathrm{mK}$ for $\mathrm{CdSe:In}$ samples that display
three dimensional Mott resistance with exponent $\zeta=1/4$ and $R(0)=6\, M\Omega\mathrm{cm}$,
significant ($\delta G/G\sim0.2$) negative magnetoresistance was
also observed in three dimensional doped n-type $\mathrm{InP}$ samples
that also show Mott law but much lower resistance $R(0)\sim10\,\Omega\mathrm{cm}$.
The paper\cite{Mitin2007} reported decrease of the resistance by
a factor of $100$ in the field $B=1T$ for $\mathrm{Ge}$ films at
$T=36\,\mathrm{mK}$ characterized by $R=400\, k\Omega$.

The complexity of the data outlined above shows that they cannot be
explained solely by a single particle theory. In particular, it cannot
explain why some materials exhibit only positive while others only
negative magnetoresistance in whole range of temperatures and magnetic
fields in the variable range hopping regime. Moreover, there are also
materials that exhibit an isotropic positive magnetoresistance only
at small fields. At larger in-plane fields the magnetoresistance of
these samples saturates, and addition of a small perpendicular field
results in a giant negative magnetoresistance \cite{Kravchenko1998,Spivak2010}.
Evidently, the spin physics plays an important role in the these materials.

Positive magnetoresistance of several orders of magnitude in high
magnetic field has been observed in many experimental works, see e.g.
\cite{Shklovskii1984,Shlimak1983,Ionov1983}. However, no data set
is sufficiently complete to allow one to associate it with the orbital
interference mechanisms \cite{Shklovskii1982} described by (\ref{eq:UniversalCor},\ref{eq:NonUniversalCor}).
For example, these works did not study the anisotropy of the magnetoresistance.

We now briefly discuss the origin of the isotropic positive magnetoresistance
in small fields which was observed in a number of works. There are
at least three possibilities. The first one is that the electron spin
polarization increases the electron energy. As a result, the density
of states at the Fermi energy changes as well. This is expected to
be a relatively small effect. An alternative mechanism associates
it with the presence of both singly and doubly occupied states near
the Fermi energy in the impurity band. In the absence of magnetic
field the process in which the electron hops from one occupied site
to another (creating a singlet) is possible. Magnetic field polarizes
spins which suppresses such processes \cite{Kamimura1985}. Thus the
magnetic field effectively changes the density of states in the impurity
band. This mechanism provides quadratic in $B$ contribution to $\log\sigma$.
Therefore, it can be effective only in the absence of the orbital
contribution, which is non-analytic in $B$.

A different mechanism might be effective if the electron system is
strongly correlated, and in the absence of disorder is close to the
Wigner-crystal-Fermi liquid transition. In the presence of disorder,
the system may be visualized as a random mixture of crystal and liquid
puddles. In this case the insulating phase corresponds to the situation
where metallic puddles do not overlap. Because the magnetic susceptibility
of the Wigner crystal is higher than that of the Fermi liquid, the
fraction of the Wigner crystal grows with increasing magnetic field
leading to the positive magnetoresistance \cite{Spivak2010}. In the
theory of $^{3}He$ this phenomenon is known as the Pomeranchuk effect.
It is possible that huge positive isotropic magnetoresistance observed
in \cite{Kravchenko1998,Gao2006,Spivak2010} in the metallic regime
of Si MOSFET's and $\mathrm{GaAs}$ quantum wells is due to this mechanism.
We believe that the same mechanism may be responsible for the positive
isotropic magnetoresistance in the hopping regime \cite{Spivak2010}.

Finally, the spin alignment in the parallel field produces the interference
between the paths and corresponds to a new mechanism of magnetoresistance.
Though this mechanism in the hopping regime has never been considered
theoretically, it is clear that it also produces a negative magnetoresistance.
We expect that this contribution will be isotropic.

While this work was in progress we learned about work\cite{Galitski2012}
that gives the arguments for the universal corrections to the magnetoresistance
of strongly disordered superconductors described by a model similar
to the electron hopping discussed here. In our terminology this model
corresponds to the case of the uniform density of states and positive
scattering amplitudes. 

\textbf{Acknowledgments.} We acknowledge useful discussions with M.
Feigelman, J. Folk, X.P.A. Gao, M. Gershenson, D. Huse, S. Kravchenko,
I. Sadovskyy, M. Sarachik and B.I. Shklovskii. B.S. thanks the International
Institute of Physics ( Natal, Brazil) for hospitality during the completion
of the paper. This research was supported by grants ARO W911NF-09-1-0395,
ANR QuDec and John Templeton Foundation. The opinions expressed in
this publication are those of the author(s) and do not necessarily
reflect the views of the John Templeton Foundation and Templeton . 

\bibliographystyle{apsrev}
\bibliography{Magnetoresistance}

\end{document}